\documentclass[
aps
prd,
preprint,
superscriptaddress,
amsfonts,
amssymb,
amsmath,
preprintnumbers,
floatfix,
nofootinbib,
unsortedaddress,
showpacs,
tightenlines,
floats,
a4paper,
eqsecnum,
]{revtex4}

\def\be{\begin{equation}}
\def\ee{\end{equation}}

\def\pa{\partial}

\def\fr{\frac}
\def\de{\delta}

\def\de{\delta}
\usepackage{color}
\usepackage{comment}

\def\vec#1{\mbox{\boldmath $#1$}} 
\def\b#1{\mbox{\boldmath $#1$}}
\usepackage{graphicx}
\usepackage[normalem]{ulem}
\newcommand{\souta}[1]{\unskip}
\newcommand*{\strike}[1]{\unskip}
\def\be{\begin{equation}}
\def\ee{\end{equation}}

\def\pa{\partial}

\def\fr{\frac}
\def\de{\delta}

\def\de{\delta}

\def\red{\textcolor{black}}
\def\cyan{\textcolor{black}}
\def\green{\textcolor{black}}
\def\magenta{\textcolor{black}}
\def\blue{\textcolor{black}}
\def\ao{\textcolor{black}}
\def\sora{\textcolor{black}}
\def\umi{\textcolor{black}}
\def\sea{\textcolor{black}}
\def\H{{\cal H}}
\def\kh{\hat{k}}
\def\f{\frac}
\newcommand{\maru}[1]{\left( #1 \right)}
\newcommand{\kaku}[1]{\left[ #1 \right]}
\newcommand{\nami}[1]{\left\{ #1 \right\}}
\newcommand{\sankaku}[1]{\left\langle #1 \right\rangle}

\makeindex

\begin{document}

\title{Primordial black holes as a 
novel probe of \\ primordial gravitational waves. II: Detailed analysis}

\author{Tomohiro Nakama}

\affiliation{Department of Physics and Astronomy, Johns Hopkins University\\ 3400 N. Charles Street, Baltimore, Maryland 21218, USA
}


\author{Teruaki Suyama}

\affiliation{Research Center for the Early Universe (RESCEU),\\
Graduate School of Science, The University of Tokyo, \\ Bunkyo-ku,
Tokyo 113-0033, Japan
}




\date{\today}

\preprint{\umi{RESCEU-19/16}}

\date{\today}

\small

\begin{abstract}
\blue{Recently we have proposed} a novel method to probe primordial gravitational waves 
\umi{from upper bounds on the abundance of} primordial black holes (PBHs). 
When the amplitude of \blue{primordial} tensor perturbations generated \ao{in the early Universe} 
\umi{is} \sora{fairly} large, 
\ao{they induce} \sora{substantial} scalar perturbations due to their second-order effects. 
If \souta{the amplitude of} \sora{these induced} scalar perturbations \sora{are} too large \sora{when they reenter the horizon}, 
then PBHs are overproduced\sora{, their abundance exceeding observational upper limits.}
\sora{That is,} \ao{primordial} tensor perturbations \ao{on superhorizon scales} \sora{can be constrained from the absence of PBHs}. 
\umi{In \souta{contrast to } our recent paper \souta{in which } we \sora{have only shown} simple estimations of \sora{these new constraints}, \sora{and hence} in this paper, we present detailed derivations, \souta{by } solving the Einstein equations for scalar perturbations induced at second order in tensor perturbations. We also derive an approximate formula for the probability density function of induced density perturbations, necessary to relate the abundance of PBHs to the primordial tensor power spectrum, 
assuming primordial tensor perturbations follow Gaussian distributions.}
\sora{Our new} \blue{upper bounds from PBHs} \sora{are compared} with other existing bounds obtained from 
big bang nucleosynthesis, cosmic microwave background, \ao{LIGO/Virgo and pulsar timing arrays.}

\end{abstract}
\maketitle
\section{Introduction}
\umi{A} \ao{stochastic \souta{gravitational wave } background \souta{(SGWB)} 
of \sora{primordial gravitational waves (PGWs) with} a huge range of wavelengths may have been generated in the early Universe}. 
The\sora{ir} power spectrum  reflects physical conditions in the early Universe, and hence \souta{constraining it or hopefully revealing at least part of it by observations} \sora{its constraints} provide valuable information for cosmology.
\sora{PGWs of} largest observable \ao{wavelengths} \sora{have} been \ao{constrained} by Planck \cite{Ade:2013zuv} and BICEP2 \cite{Ade:2014xna}, 
\souta{\umi{Components of} SGWB}
\sora{while those of}\ao{\souta{with} shorter wavelengths} \umi{have been} constrained 
\souta{by inferring}
\sora{by limits on} \souta{the value of} $N_{\rm{eff}}$, 
the effective number of degrees of freedom of relativistic \ao{fermions}, 
at big bang nucleosynthesis (BBN) through the current abundance of \sora{the} light elements \cite{Allen:1996vm}, 
or at photon decoupling through the anisotropy of cosmic microwave background (CMB) \cite{Smith:2006nka,Kikuta:2014eja}. 
\ao{Recently} \sora{PGWs} on small\sora{er} scales \sora{have been}\souta{been argued to be also} constrained by \souta{the non-detection of} \sora{upper limits on the }deviation of the CMB photons' energy spectrum from the Planck distribution
\cite{Ota:2014hha,Chluba:2014qia}. 
\sora{Though }BBN and CMB \ao{constrain \sora{PGWs} of a \sora{wide} range of wavelengths, \souta{ while
each ground-based laser interferometer \ao{constrains SGWB} on a relatively limited frequency range. 
However, }}
\ao{\sora{these }upper bounds}, obtained through $N_{\rm{eff}}$, \umi{entail} an assumption about 
the number of relativistic species in the early Universe, as is discussed later. 
\ao{Furthermore, to obtain} BBN or CMB bounds we implicitly assume that 
any physical mechanisms, both known and unknown, \textit{increase} $N_{\rm{eff}}$, 
\ao{from} the standard value $N_{\rm{eff}}=3.046$ \cite{Mangano:2005cc}. 
However, $N_{\rm{eff}}$ can decrease e.g. in brane world scenarios  \cite{Ichiki:2002eh,Apostolopoulos:2005at,Maartens:2010ar}.
\souta{\ao{It} would be desirable to have another independent cosmological method to probe 
SGWB on a wide range of scales, which does not depend on the assumptions mentioned above \ao{much}, }
\sora{Recently we} \umi{proposed} \souta{such} a \sora{new} method \sea{to constrain PGWs} in
our recent work \cite{Nakama:2015nea}, 
\sora{which is also applicable on a wide range of wavelengths and in addition does not depend on the aforementioned assumptions much.}
\umi{In this paper, we present \souta{the } detailed derivation\sora{s} of the results presented \souta{in \cite{Nakama:2015nea}} \sora{there}.}

\souta{\ao{\umi{In order to constrain SGWB we} consider}}\sora{Our new method uses} the formation of 
primordial black holes (PBHs), \umi{formed} in the early Universe, \strike{\sora{that is,}} well before the \blue{cosmic} structure formation. 
\ao{One} of the simple and \ao{plausible} mechanisms \ao{to form PBHs} is the direct collapse of \ao{density fluctuations} 
during \ao{the radiation-dominated era}, which \ao{\souta{is triggered}\sora{happens} when the fractional density perturbation} 
\sora{of} order unity 
\souta{at the moment of the horizon crossing of the perturbation}
\sora{reenters the \sea{Hubble} horizon} \cite{Zel'dovich-1974,Carr:1974nx,Carr:1975qj}. 
See also \cite{Harada:2013epa} for an updated discussion of the formation condition and \souta{see also} \cite{1978SvA....22..129N,Shibata:1999zs,Niemeyer:1999ak,Polnarev:2006aa,Polnarev:2012bi,Nakama:2013ica,Nakama:2014fra} for numerical simulations of the \sea{PBH} formation \souta{of PBHs} \sora{process}.
There is no conclusive evidence for the existence of PBHs \ao{in the present as well as in the past} and upper bounds on the\sora{ir} 
abundance \souta{of PBHs } \strike{\ao{o\sora{n} various masses}} 
\sea{over a wide mass range} have been obtained by various \ao{methods} (see e.g. \cite{Carr:2009jm} and references therein). 
\umi{One of the cosmological implications of the\sora{ir} absences \souta{of PBHs }is to constrain} the power spectrum of the curvature perturbation  \cite{Bugaev:2008gw,Josan:2009qn}\footnote{
\blue{Other methods to constrain primordial \umi{scalar} perturbations on small scales include CMB spectral distortions 
\cite{Barrow2,Chluba:2011hw,Chluba:2012we,Chluba:2012gq,Dent:2012ne,Khatri:2012tw,Sunyaev:2013aoa,Khatri:2013dha,Chluba:2013dna,Chluba:2013pya}, 
acoustic reheating \cite{Nakama:2014vla,Jeong:2014gna} \sora{and} ultracompact minihalos \cite{Bringmann:2011ut,Kohri:2014lza}.
}
}.
\ao{\umi{In a broader context, PBHs} provide valuable information to} exclude models of the early Universe which predict \sora{an overproduction of \strike{them}} \sea{PBHs} \souta{too many PBHs}.

\blue{As we have briefly discussed in our recent work \cite{Nakama:2015nea},
PBHs can also be used to constrain \ao{tensor perturbations generated in the early Universe\sora{, exiting the horizon once} and reentering the horizon \umi{later\souta{ at some moment in time in the history of the Universe}}.}} 
\ao{This} is because large \souta{amplitude }tensor perturbations induce \blue{large} scalar perturbations (\textit{induced scalar perturbations}) \ao{at second order in tensor perturbations}. 
\ao{If \souta{the amplitude of} primordial tensor perturbations \sora{are} too large, \souta{that of} \umi{\sora{\strike{these resulting} }induced scalar} perturbations become\souta{s} also too large, and then they collapse \umi{to overproduce PBHs shortly after} their horizon reenty, exceeding \souta{observational} \sora{existing upper} limits\souta{ \umi{on them}}.}
\sora{That is, }\souta{In other words,} \souta{the amplitude of }\ao{primordial tensor perturbations can be constrained} \sora{from upper limits on}\souta{requiring } PBHs\souta{ are not overproduced}\footnote{\label{2}
\souta{In several papers, s}Second-order effects of scalar perturbations to induce tensor perturbations (termed induced gravitational waves) have 
been discussed \sora{in the literature} \cite{Matarrese:1993zf,Matarrese:1997ay,Carbone:2004iv,Ananda:2006af,Baumann:2007zm,Alabidi:2012ex,Bugaev:2010bb}; we can place upper bounds on \souta{the amplitude of } scalar perturbations(, which can be translated into upper bounds on the 
abundance of PBHs \cite{Saito:2008jc,Saito:2009jt,Bugaev:2009zh},)
from the non-detection of GWs. 
\sora{Note that }our present paper \sora{discusses an effect} opposite to \sora{this generation of induced gravitational waves}. 
}\footnote{
The direct gravitational collapse of nonlinear localized gravitational waves has been discussed in the literature 
\cite{Brill:1959zz,Eppley:1977dk,Miyama:1981mh,Shibata:1993fx,Shibata:1995we,Anninos:1996fg,Shibata:1997ix,Alcubierre:2000xu,Pfeiffer:2004qz} 
and so \ao{tensor perturbations may \sora{also} be constrained} using this \souta{direct collapse mechanism} \sora{phenomenon}. 
\souta{However} \sora{Still}, the initial conditions and dynamics of \sora{cosmological} nonlinear gravitational waves \souta{originated from SGWB } during the \ao{radiation-dominated era} 
have not been \umi{well} understood\sea{. Since the dynamics of nonlinear radiation density perturbations is better understood, we consider only scalar perturbations induced by the tensor perturbation.}
\strike{and so in this paper we consider \souta{the } \ao{second-order tensor perturbations \souta{to induce} \sora{inducing} scalar perturbations}, 
\sora{since} the dynamics of nonlinear radiation density perturbations is better understood. }
}. 
\blue{Whereas we have presented only simple estimations to 
obtain \souta{upper bounds from PBHs} \sora{these new constraints} in \cite{Nakama:2015nea}, 
\sora{in the present paper} we show detailed \souta{analysis} \sora{derivations} \souta{to obtain } \sora{for them}.}

\color{black}
Due to our ignorance of the  physics \umi{in} the early Universe, 
new upper limits on tensor perturbations on small scales in themselves would be worthwhile. 
In addition, 
there are models of the early Universe 
\sea{\cite{Boyle:2003km,Baldi:2005gk,Copeland:2008kz,Kobayashi:2011nu,Biswas:2014kva,Ashoorioon:2014nta,Cannone:2014uqa,Graef:2015ova}}
which 
can predict large tensor perturbations on small scales, 
\ao{which makes our new upper limits even more valuable \sea{(see the next section)}.}  
\sora{Note that, }\ao{if} a model predicts large tensor perturbations on small scales \umi{but also large} scalar perturbations at the same time, 
then such a model would be more severely constrained \ao{\umi{from} the absence of PBHs \souta{since additional PBHs are } generated from the first-order scalar perturbations.}
\umi{In this paper} we consider PBH formation only from \sea{induced scalar perturbations,} second order \sea{in} tensor perturbations, \ao{and thus}
our bounds on tensor perturbations are conservative or model-independent, in the sense that th\sora{e}se bounds do not depend on \souta{the amplitude of }first-order scalar perturbations on small scales. 
\ao{Importantly,} there are models of the early Universe which predict not only large tensor perturbations\strike{ on small scales}, but also large \textit{tensor-to-scalar ratio} on small scales, and our PBH bounds are particularly 
useful to constrain these types of models, some of which are  \umi{reviewed} \ao{in the next section}.

\color{black}

This paper is organized as follows; 
\ao{In Sec. II we \umi{review} some of the early Universe models 
\sora{\cite{Boyle:2003km,Baldi:2005gk,Copeland:2008kz,Kobayashi:2011nu,Biswas:2014kva,Ashoorioon:2014nta,Cannone:2014uqa,Graef:2015ova}} which predict large tensor-to-scalar ratio on small scale\sora{s}.}
In Sec. III the radiation density perturbation generated from \strike{\sora{second-order}} tensor perturbations is calculated. 
Section IV is dedicated to \souta{the} \sora{a} discussion of upper bounds on tensor modes from PBHs along with \souta{the} \sora{a} comparison \souta{of them } with 
those obtained from \umi{other methods, 
and} we conclude in Sec. V. 
\ao{
\section{Early Universe models predicting large tensor-to-scalar ratio on small scales}
}
In \cite{Boyle:2003km}, \strike{blue} tensor power spectra \sea{were shown to be blue} (i.e. larger power on smaller scales)
\strike{were obtained} in cyclic/ekpyrotic models, with 
the spectrum of scalar perturbations \souta{being } kept slightly red (smaller power on smaller scales) to match observations on large scales. The cyclic Universe entails 
the periodic collisions of orbifold planes moving 
in an extra spatial dimension, which is equivalently 
described by a scalar field rolling back and forth in an effective potential. 
Each cycle consists of an accelerated expansion phase, a slow contraction phase (the ekpyrotic phase), during which the Universe is dominated by the kinetic energy as well as \sora{the} negative potential energy of the scalar field and primordial fluctuations are generated, a rapid contraction phase followed by a bounce at which matter and radiation are generated, a phase 
dominated by the kinetic energy of the scalar field, 
a radiation\sora{-}dominated, expanding phase, 
and finally \sora{a} \souta{matter\sora{-} and dark\sora{-}energy\sora{-dominated}} phase \sora{dominated by matter and dark energy}. 
The spectrum of scalar perturbations can be adjusted to be slightly red by \souta{the choice of} \sora{tuning} the scalar field potential during the ekpyrotic phase, and the tensor spectrum turns out to be blue up to the scale corresponding to the end of the ekpyrotic phase. For early Universe scenarios where the spectrum of tensor perturbations is strongly blue, 
probing them on CMB scales may be challenging, while constraints on small-scale components, 
such as those discussed in this \umi{paper}, 
may provide useful information.
Indeed, they noted that the strongest constraint on \umi{their} model parameters is obtained from BBN constraints on high-frequency \sora{P}GWs. 

If the inflaton violates the null energy condition (NEC, $\rho+p\geq 0$), the Hubble parameter increases during inflation (\ao{super inflation}) and the spectral tilt 
$n_T$ becomes positive, since $n_T=-2\epsilon\equiv 2\dot{H}/H^2\propto -(\rho+p)$.\strike{ 
They} \sea{In \cite{Baldi:2005gk} it was shown} \strike{pointed out} that \souta{it is possible 
to violate } NEC \sora{can be violated} without the instability of 
fluctuations of the inflaton. 
\strike{They introduced a toy model}
\sea{There a toy model was introduced}, 
with the energy density of the NEC-violating inflaton  $\rho=\ao{-\dot{\phi}^2/2+}V_0e^{-\lambda\phi/M_{\mathrm{pl}}}$, which leads to a stage of pole-like inflation, when $a(t)\sim(-t)^p,\: t<0,\: p=-2/\lambda^2<0$. The background and fluctuations  are shown to be stable at the classical level. 
It was noted that in this model, some mechanism, quantum effects or another field, is necessary to avoid singularity at $t\rightarrow 0$ and \umi
{to}
drive the Universe into a radiation\sora{-}dominated epoch. 

The spectrum of tensor perturbations generated 
during \sora{a} super inflation in the framework of 
loop quantum cosmology (LQC) is calculated in 
\cite{Copeland:2008kz}. There a strong blue 
tile with $n_T\simeq 2$ was obtained, 
while the form of the inflaton potential to realize 
a scale-invariant power spectrum of scalar perturbations was also discussed 
in their previous works. 
In their scenario, the nondimensional power spectrum of tensor perturbations 
on smallest scales is roughly given by the square of the Hubble parameter $H_{\rm{e}}$ at the end of inflation 
in units of the Planck scale, and this \sora{implies that} $H_{\rm{e}}$ can be constrained \sora{e.g.} by our PBH constraints\souta{ discussed \sora{in the present paper}}. 
They note that $H_{\rm{e}}$ is, in principle, \sora{also} related to the amplitude of scale-invariant curvature perturbations as well, but such a relation has not been obtained yet in the scenarios they consider.

Large tensor perturbations on small scales may also be realized in the framework of the so-called generalized G-inflation ($G^2$-inflation) \cite{Kobayashi:2011nu}. 
The action of $G^2$-inflation
contains 
four generic functions $K, G_3, G_4, G_5$ of $\phi$ and 
$X=-\partial_\mu\phi\partial^\mu\phi/2$. The quadratic action for the tensor perturbations is 
\begin{equation}
S_T^{(2)}=\frac{1}{8}\int dtd^3xa^3
\left[{\cal G}_T\dot{h}_{ij}^2-\frac{{\cal F}_T}{a^2}(\nabla h_{ij})^2\right],
\end{equation}
\begin{equation}
{\cal G}_T\equiv 2\left[
G_4-2XG_{4X}-X(H\dot{\phi}G_{5X}-G_{5\phi})
\right],\quad
{\cal F}_T\equiv 2\left[G_4-X(\ddot{\phi}G_{5X}+G_{5\phi})\right].
\end{equation}
The squared sound speed is $c_T^2={\cal F}_T/{\cal G}_T$, 
which is not necessarily unity in general cases. 
The parameters $\epsilon\equiv -\dot{H}/H^2$, $f_T\equiv 
\dot{{\cal F}_T}/H{\cal F}_T$ and $g_T\equiv \dot{{\cal G}_T}/H{\cal G}_T$ are introduced and they are assumed to be nearly constant.
The nondimensional power spectrum of the tensor perturbations was obtained as 
\begin{equation}
{\cal P}_T=8\gamma_T\frac{{\cal G}_T^{1/2}}{{\cal F}_T^{3/2}}\frac{H^2}{4\pi^2}\bigg|_{-ky_T=1},
\end{equation}
where 
\begin{equation}
\nu_T\equiv \frac{3-\epsilon+g_T}{2-2\epsilon-f_T+g_T},
\quad \gamma_T=2^{2\nu_T-3}\bigg|\frac{\Gamma(\nu_T)}{\Gamma(3/2)}\bigg|^2\left(1-\epsilon-\frac{f_T}{2}+\frac{g_T}{2}\right),
\quad dy_T\equiv \frac{c_T}{a}dt.
\end{equation}
The tensor spectral tilt is given by $n_T=3-2\nu_T$, 
and the tensor spectrum is blue ($0<n_T$) if $4\epsilon+3f_T-g_T<0.$ Also, if the sound speed becomes temporarily small, \souta{the amplitude of } tensor perturbations 
\souta{is} \sora{are} enhanced on the corresponding scales. 


A slightly red spectrum of the curvature perturbation, while keeping the \souta{gravitational} \sora{tensor} 
spectrum strongly blue-tilted, was also shown to be realized during a stringy thermal 
contracting phase at temperatures beyond the so-called Hagedorn temperature (the Hagedorn phase) in \cite{Biswas:2014kva}, assuming a 
nonsingular bounce. 
In that scenario, primordial curvature perturbations originate from statistical thermal fluctuations, 
not by scalar field quantum fluctuations. 

Scalar and tensor perturbations in large field chaotic models with non-Bunch-Davies (non-BD) initial states were analyzed in \cite{Ashoorioon:2014nta}, and it was shown that in that model also gravitational waves can be blue while maintaining slightly red scalar perturbations.
Normally, initial states for perturbations are chosen to be Bunch-Davies (BD) vacuum states, 
namely, perturbation modes on sub-Hubble scales effectively propagate in vacuum states associated with flat space. Non-BD initial states were characterized by the Bogoliubov coefficients for each $k$ mode and for both scalar and tensor perturbations, which were denoted by $\alpha_k^S,\beta_k^S,\alpha_k^T,\beta_k^T,$
with $(\alpha_k^{S,T},\beta_k^{S,T})=(1,0)$ 
corresponding to the standard BD initial states. 
These parameters are determined by unknown high energy physics, 
and depending on the choice of the above parameters, blue gravitational waves were obtained while maintaining the scalar perturbations slightly red. 

Blue \ao{gravitational} waves with slightly red scalar perturbations were also obtained without violating NEC by breaking the spatial diffeomorphism, usually imposed on the dynamics of perturbation\sora{s}, in the context of effective theory of inflation 
\cite{Cannone:2014uqa,Graef:2015ova}. 
There, breaking of spatial diffeomorphism was considered by effective quadratic mass terms or derivative operators for metric fluctuations in the Lagrangian during inflation without the necessity for specifying the UV completion, while noting that it may be a version of massive gravity coupled to an inflaton, some model of inflation using vectors, or sets of scalars obeying some symmetries.

\sea{Before closing this section, let us emphasize one important assumption made throughout this paper.}
\sora{\strike{In this paper we}\sea{We} calculate evolution of primordial fluctuations assuming \strike{the standard cosmology described by} \sea{they obey} general relativity below some energy scale\strike{, though large primordial tensor-to-scalar ratio may be realized by e.g. some modification of gravity, a few examples of which are mentioned above}. That energy scale and comoving \strike{wave number} \sea{wave number} $k$ of primordial fluctuations are related as follows. The wave number $k$ is said to reenter the horizon when $k=aH$, where $a$ and $H$ are the scale factor and the Hubble parameter. The scale factor can be eliminated by the relation $H^2=H_0^2\Omega_ra^{-4}$, where $\Omega_r$ is the radiation density parameter and $H_0$ is the current Hubble parameter and here they are taken as $\Omega_r=5\times 10^{-5}$ and $H_0=67\mathrm{km/s/Mpc}$.
The Hubble parameter $H$ and the temperature of the Universe $T$ are related by (in natural units)  $H^2=4\pi^3g_*T^4/45,$ where $g_*$ is the degrees of freedom of relativistic species here taken as $g_*=106.75$. From these relations the temperature and comoving \strike{wave number} \sea{wave number} are related by 
\begin{equation}
T=\left(\frac{4\pi^3}{45}g_*\right)^{-\frac{1}{4}}(H_0\Omega_r^{1/2}k^{-2})^{-\frac{1}{2}}\simeq 5\times 10^{10}\mathrm{GeV}\left(\frac{k}{10^{18}\mathrm{Mpc}^{-1}}\right).
\end{equation}
For instance, if the theory is reduced to the standard cosmology described by general relativity at $T=5\times 10^{10}$GeV, then our upper limits summarized in Fig. \ref{constraint} are applicable for $k<10^{18}\mathrm{Mpc}^{-1}$.
}
\clearpage
\umi{
\section{Radiation density perturbations generated from \\tensor perturbations}
}
We work in the comoving gauge, \sora{in which} the metric is written as\footnote{
\green{
Perturbations to the metric and energy momentum tensor are written as (see \cite{Weinberg} for more details)
\begin{equation}
ds^2=a^2[-(1+2\Phi)d\eta^2\sora{+}2B_{,i}d\eta dx^i+\nami{(1-2\Psi)\delta_{ij}-2E_{,ij}-2h_{ij}}dx^idx^j],
\end{equation}
\begin{equation}
T_{\mu\nu}=(p+\delta p)g_{\mu\nu}+(\rho+\delta \rho+p+\delta p)(u_\mu+\delta u_\mu)(u_\nu+\delta u_\nu),
\end{equation}
where the \sora{spatial components of the} velocity perturbation $\delta u_\mu$ \sora{are} written as $\delta u_i=\delta u_{,i}$.
Let us consider a coordinate transformation of the form $x^\mu\rightarrow x^\mu+\epsilon^\mu(x^\mu)$, with 
$\epsilon_0=-\epsilon^0,\; \epsilon_i=a^2\epsilon^i,$
$\epsilon_i=\epsilon_{,i}.$
Then $E$ and $\delta u$ transform as 
$E\rightarrow E+\epsilon\sora{/a^2},$
$\delta u\rightarrow \delta u-\epsilon_0$.
Here we choose $\epsilon$ so that $E=0$, and then choose $\epsilon_0$ so that $\delta u=0$. 
Both choices are unique, so that there is no freedom to make \blue{further} gauge transformations. 
\sora{This choice is sometimes called the comoving gauge (e.g. \cite{Baumann:2009ds}). }
}
} 
\be
ds^2=a^2[
-(1+2\Phi)d\eta^2\green{+}2B_{,i}d\eta dx^i+((1-2\Psi)\delta_{ij}\green{+}2h_{ij})dx^idx^j
],\label{metric}
\ee 
\ao{where $h_{ij}$ is the tensor perturbation satisfying $h_{ij,i}=h_{ii}=0$.
Throughout} this paper it is assumed that the amplitude of initial tensor perturbations 
is much larger than that of scalar perturbations (schematically, $(\rm{scalar})\ll (\rm{tensor})$), and 
so the scalar quantities in the metric above should be regarded as second order in $h_{ij}$. 
Hence, for scalar perturbations we write down the Einstein equations 
keeping second-order terms only in $h_{ij}$. 
\umi{As is also mentioned in the Introduction}, our \blue{upper bounds from PBHs} on tensor perturbations thus obtained are 
applicable even if this initial hierarchy between tensor and scalar perturbations 
\umi{does not hold}. This is because if the amplitude of 
scalar perturbations is as larger as, or larger than that of tensor perturbations, 
then the abundance of PBHs increases when the amplitude of tensor modes is fixed. 
Namely, assuming $(\rm{scalar})\ll (\rm{tensor})$ initially is most conservative 
in placing upper bounds on tensor modes, and hence 
our bounds are applicable even if that assumption does not hold. 

Let us write down the fundamental equations in the following. 
We denote the energy density and pressure of \sora{the} dominating radiation 
by $\rho$ and $p$, respectively, and write $p=c_{\rm{s}}^2\rho$, where $c_{\rm{s}}$ is the speed of sound. 
In this paper we restrict our attention to the formation of PBHs due to collapse of radiation density perturbations during \umi{the radiation-dominated era,}
and so we set $c_{\rm{s}}=1/\sqrt{3}$ in calculations, though we leave $c_{\rm{s}}$ unspecified in equations below for generality. 
We decompose $\rho$ and $p$ as $\rho(\eta,\b{x})=\rho_0(\eta)+\delta\rho(\eta,\vec{x})$ and $p(\eta,\vec{x})=p_0(\eta)+\delta p(\eta,\vec{x})$.

The zeroth-order Einstein equations yield
\be
{\cal H}^2=\fr{8\pi G}{3}a^2\rho_0,\label{zero1}
\ee
\be
{\cal H}^2-{\cal H}'=4\pi Ga^2(\rho_0+p_0),\label{zero2}
\ee
where ${\cal H}\equiv a'/a$ with \sora{the} prime denoting differentiation with respect to the conformal time $\eta$.
These two equations are combined to give 
\be
2\H'+(1+3c_{\rm{s}}^2)\H^2=0.\label{background}
\ee

The Einstein equations at first order \umi{in $h_{ij}$ give} the standard evolution equation for tensor modes as follows:
\be
h_{ij}''+2\H h_{ij}'-\Delta h_{ij}=0.\label{linear}
\ee 

The Einstein equations at second order \blue{in $h_{ij}$}, derived in Appendix A, are as follows:
\be
\Delta\Psi-3{\cal H}(\Psi'+{\cal H}\Phi)-\H\Delta B
+S_1
=4\pi G a^2\delta \rho,\label{timetime}
\ee
\be
(\Psi'+\H \Phi+S_2)_{,i}=0,\label{timespace}
\ee
\be
\Psi''+{\cal H}(2\Psi+\Phi)'+(2{\cal H}'+\H^2)\Phi
+\fr{1}{2}\Delta(\Phi-\Psi+B'+2\H B)
+S_3+S_4
=4\pi G a^2\delta p, \label{ein3}
\ee
\be
(\Phi-\Psi+B'+2\H B-2S_5)_{,ij}=0.\label{ein4}
\ee
\umi{In these equations} the following terms, second order in $h_{ij}$, source the scalar perturbations:
\be
S_1\equiv 
-\fr{1}{4}h_{ij}'h^{ij'}-2\H h_{ij}h^{ij'}
+h_{ij}\Delta h^{ij}-\fr{1}{2}\pa_jh_{ik}\pa^{k}h^{ij}
+\fr{3}{4}\pa_kh_{ij}\pa^{k}h^{ij},\label{S10}
\ee
\umi{
\be
\Delta S_2=\pa^iS_i,
\quad S_i=-h^{jk}\pa_kh_{ij}'+\fr{1}{2}h^{jk'}\pa_ih_{jk}+h^{jk}\pa_ih_{jk}',
\ee
}
\be
S_{3}\equiv
\fr{3}{4}h_{ij}'h^{ij'}+h_{ij}h^{ij''}+2\H h_{ij}h^{ij'}
-h_{ij}\Delta h^{ij}+\fr{1}{2}\pa_jh_{ik}\pa^{k}h^{ij}
-\fr{3}{4}\pa_kh_{ij}\pa^{k}h^{ij},
\ee
\be
\Delta S_4=\fr{1}{2}(\Delta S^i_{~i}-\pa^i\pa^jS_{ij}),\label{S4}
\ee
\be
\Delta^2S_5=\fr{1}{2}(3\pa^i\pa^jS_{ij}-\Delta S^i_i),\label{S5}
\ee
\begin{align}
S_{ij}\equiv
-h_i^{~k'}h_{jk}'-h_{ik}h_{j}^{~k''}-2\H h_{i}^{~k}h_{jk}'
+h^{kl}\pa_k\pa_lh_{ij}+h_{i}^{~k}\Delta h_{jk}
-h^{kl}\pa_l\pa_ih_{jk}-h^{kl}\pa_l\pa_jh_{ik}\nonumber\\
-\pa_kh_{jl}\pa^lh_{i}^{~k}+\pa_lh_{jk}\pa^{l}h_{i}^{~k}
+\fr{1}{2}\pa_ih_{kl}\pa_jh^{kl}
+h^{kl}\pa_i\pa_jh_{kl}.
\end{align}
Using (\ref{linear}),
\umi{$S_1, S_3$ and $S_{ij}$} are rewritten as follows:
\be
S_1=
-\fr{1}{4}h_{ij}'h^{ij'}
+h_{ij}h^{ij''}-\fr{1}{2}\pa_jh_{ik}\pa^{k}h^{ij}
+\fr{3}{4}\pa_kh_{ij}\pa^{k}h^{ij},
\ee
\be
S_{3}=
\fr{3}{4}h_{ij}'h^{ij'}+\fr{1}{2}\pa_jh_{ik}\pa^{k}h^{ij}
-\fr{3}{4}\pa_kh_{ij}\pa^{k}h^{ij},
\ee
\begin{align}
S_{ij}=
-h_i^{~k'}h_{jk}'
+h^{kl}\pa_k\pa_lh_{ij}
-h^{kl}\pa_l\pa_ih_{jk}-h^{kl}\pa_l\pa_jh_{ik}\nonumber\\
-\pa_kh_{jl}\pa^lh_{i}^{~k}+\pa_lh_{jk}\pa^{l}h_{i}^{~k}
+\fr{1}{2}\pa_ih_{kl}\pa_jh^{kl}
+h^{kl}\pa_i\pa_jh_{kl}.
\end{align}
The conservation of the energy-momentum tensor yields
\be
\de\rho'+3\H(\de\rho+\de p)-(\rho+p)\Delta B
-3(\rho+p)\Psi'-2(\rho+p)h^{ij}h_{ij}'=0,\label{cons1}
\ee
\be
\pa_i(\de p
+(\rho+p)\Phi)=0.\label{cons2}
\ee

\green{
\umi{From these equations} one can derive \sora{the} evolution equation of $\Psi$ as follows.
First, Eqs. (\ref{cons1}) and (\ref{cons2}) lead to (hereafter we work in Fourier space)
\begin{equation}
\Phi'=-c_s^2\maru{-k^2B+3\Psi'+2h^{ij}h_{ij}'}.
\end{equation}
The term $-k^2B$ of the above can be eliminated by the following relation, 
obtained from Eqs. (\ref{timetime}) and (\ref{timespace}):
\begin{equation}
-k^2B=\f{-k^2\Psi}{{\cal H}}+3S_2+\f{S_1}{\cal H}-\f{3}{2}{\cal H} \delta_r,
\end{equation}
where $\delta_r\equiv \delta\rho/\rho_0$.
Using these and (\ref{ein4}) as well as (\ref{background}), (\ref{ein3}) can be rewritten as
}
\be
\Psi''+2\H\Psi'+c_{\rm{s}}^2k^2\Psi=S.\label{evolution}
\ee
Here,  
\be
S\equiv c_{\rm{s}}^2S_1-S_3-\kh^i\kh^jS_{ij}+2c_{\rm{s}}^2\H h^{ij}h_{ij}'\label{source0}
\ee
is the source term representing generation of scalar perturbations due to the \strike{second-order} tensor perturbations. 
From (\ref{timespace}) and (\ref{cons2}), the energy density perturbation is given by 
\be
\delta_r
=\fr{1+c_{\rm{s}}^2}{c_s^2\H}(\Psi'+S_2).\label{density}
\ee

Eq. (\ref{evolution}) can be formally solved as\footnote{
We choose $\eta=0$ at the beginning of the radiation-dominated era, and 
we assume the initial condition is $\Psi(0,\vec{k})=0$. Strictly speaking, however, $\Psi$ is also generated before the radiation-dominated era at second order in tensor perturbations, even without intrinsic first-order scalar perturbations. That generation is highly model-dependent, and hence we restrict attention to the generation of $\Psi$ only during the radiation-dominated era to adopt the above initial condition. This neglect of the generation of $\Psi$ before the radiation-dominated era would probably lead to conservative upper bounds on tensor perturbations, since in general $\Psi$ would be larger if the generation before $\eta=0$ is additionally taken into account. 
An analogous assumption is also made in the literature
discussing induced gravitational waves (see footnote\ref{2}).
}
\be
\Psi(\eta,\vec{k})=a^{-1}(\eta)\int_0^\eta d\tilde{\eta}g_k(\eta,\tilde{\eta})a(\tilde{\eta})S(\tilde{\eta},\vec{k}),
\label{solution}
\ee
where $g_k$ is the \blue{retarded} Green's function satisfying
\be
g_k''+\left(c_{\rm{s}}^2k^2-\fr{a''}{a}\right)g_k=\delta(\eta-\tilde{\eta}).
\ee
\magenta{
\sora{During the radiation-dominated epoch, its} \umi{ solution} can be constructed by the two homogeneous solutions 
\be
v_1(k,\eta)=\sin(c_{\rm{s}}k\eta),\quad v_2(k,\eta)=\cos(c_{\rm{s}}k\eta)
\ee 
as follows \cite{Baumann:2007zm}:
\be
g_k(\eta,\tilde{\eta})
=\fr{v_1(k,\eta)v_2(k,\tilde{\eta})-v_1(k,\tilde{\eta})v_2(k,\eta)}{v_1'(k,\tilde{\eta})v_2(k,\tilde{\eta})-v_1(k,\tilde{\eta})v_2'(k,\tilde{\eta})}
=\fr{1}{c_{\rm{s}}k}\sin\left(c_{\rm{s}}k(\eta-\tilde{\eta})\right) \:\:\blue{\mathrm{for}\:\: \eta\geq \tilde{\eta}\:.}
\ee
}

The two point correlation function of $\Psi$ can be expressed as, \blue{denoting its nondimensional power spectrum by ${\cal P}_\Psi$},
\begin{align}
&\langle\Psi(\eta,\vec{k})\Psi^{*}(\eta,\vec{K})\rangle
=\fr{2\pi^2}{k^3}\delta(\vec{k}-\vec{K}){\cal P}_\Psi(k)\nonumber\\
&=a^{-2}(\eta)\int_{0}^{\eta}d\eta_1\int_{0}^{\eta}d\eta_2g_k(\eta,\eta_1)g_K(\eta,\eta_2)a(\eta_1)a(\eta_2)\langle S(\eta_1,\vec{k})S(\eta_2,\vec{K})\rangle.
\end{align}
In the following, let us write down the Fourier components of the source \umi{$S$, given by } \blue{(\ref{source0})}. 
\umi{We begin by decomposing} $h_{ij}(\eta,\vec{x})$ \umi{as} (following \cite{Saito:2009jt}):
\be
h_{ij}(\eta,\vec{x})=\int\fr{d^3\vec{k}}{(2\pi)^{3/2}}e^{i\vec{k}\cdot\vec{x}}
(h^{+}(\eta,\vec{k})e^{+}_{ij}(\vec{k})+h^{\times}(\eta,\vec{k})e^{\times}_{ij}(\vec{k})),
\ee
where for $\vec{k}$ in the z-direction
\be
e^{+}_{11}(\hat{z})=-e^{+}_{22}(\hat{z})=e^{\times}_{12}(\hat{z})=e^{\times}_{21}(\hat{z})=1, \quad \mathrm{others}=0
\ee
while for $\hat{k}\sora{\equiv \vec{k}/|\vec{k}|}$ in any other direction, $e^r_{ij}(\hat{k})(r=+,\times)$ is defined by applying 
on each of the indices $i$ and $j$ a standard rotation, that takes the z-direction into 
the direction of $\hat{k}$ (see e.g. \cite{Weinberg}). 
Then one can check the following: 
\be
\displaystyle \sum_{ij}e^{r}_{ij}(\vec{k})e^{s}_{ij}(\cyan{\vec{k}})=2\delta^{rs}.
\ee 
Let us further decompose \umi{the} Fourier components as $h^r(\eta,\vec{k})=D(\eta,k)h^r(\vec{k})$, 
where $h^r(\vec{k})$ is the initial amplitude and $D(\eta,k)$ is the growth factor, which can be 
obtained by solving the linear evolution equation (\ref{linear}) for $h_{ij}$ \umi{(dropping the decaying mode)}:
\be
D(\eta,k)=\frac{\sin k\eta}{k\eta}.
\ee
It turns out that \umi{the Fourier components of the source $S$} can be written as follows (see Appendix \blue{B}):
\be
S(\eta,\vec{k})=\sum_{rs}\int\fr{d^3\vec{k}'}{(2\pi)^{3/2}}h^r(\vec{k}')h^s(\vec{k}-\vec{k}')
A_{rs}(\eta,\vec{k},\vec{k}'),\label{source}
\ee
\be
A_{rs}(\eta,\vec{k},\vec{k}')
\equiv
f_1(\eta,\vec{k},\vec{k}')E_1^{rs}(\vec{k},\vec{k}')
+f_2(\eta,\vec{k},\vec{k}')E_2^{rs}(\vec{k},\vec{k}').\label{Ars}
\ee
Here,
\be
\blue{
E_1^{rs}(\vec{k},\vec{k}')\equiv \kh_j\kh^ke_{ik}^r(\vec{k}')e^{ij}_s(\vec{k}-\vec{k}'),\quad E_2^{rs}(\vec{k},\vec{k}')\equiv e_r^{ij}(\vec{k}')e^s_{ij}(\vec{k}-\vec{k}'),
}
\ee
and their nonzero components are written as\footnote{
These expressions are obtained by first setting $\hat{k}=\hat{z}$, which is possible due to isotropy, 
and by assuming $\hat{k'}$ is on the $\blue{z}-y$ plane, which is justified by the rotational invariance of 
$E_1^{rs}$ and $E_2^{rs}$. 
}
\begin{equation}
E_1^{++}(\vec{k},\vec{k}')=-\mu_1\sqrt{1-\mu^2}\sqrt{1-\mu_2^2},\quad
E_1^{\times\times}(\vec{k},\vec{k}')
=-\sqrt{1-\mu^2}\sqrt{1-\mu_2^2},\label{E1}
\end{equation}
\begin{equation}
E_2^{++}(\vec{k},\vec{k}')=1+\mu_1^2,\quad E_2^{\times\times}(\vec{k},\vec{k}')=2\mu_1,\label{E2}
\end{equation}
where $\mu\equiv \vec{k}\cdot\vec{k}'/k\,k'$ and
\begin{equation}
\mu_1\equiv\fr{\vec{k}'\cdot(\vec{k}-\vec{k}')}{k'|\vec{k}-\vec{k}'|}=\fr{k\mu-k'}{|\vec{k}-\vec{k}'|},\quad
\mu_2\equiv\fr{\vec{k}\cdot(\vec{k}-\vec{k}')}{k|\vec{k}-\vec{k}'|}=\fr{k-k'\mu}{|\vec{k}-\vec{k}'|}.
\end{equation}
\blue{
Also the above $f_1$ and $f_2$ are given by (see Appendix B)
}
\be
f_1(\eta,\vec{k},\vec{k}')=D(\eta,k')\left\{
\overleftarrow{\pa_\eta}\pa_\eta-\fr{1}{2}(3-c_{\rm{s}}^2)k^2+3kk'\mu-k^{'2}
\right\}D(\eta,|\vec{k}-\vec{k}'|),\label{f1}
\ee
\begin{align}
f_2(\eta,\vec{k},\vec{k}')=D(\eta,k')\left\{
-\fr{1}{4}(3+c_{\rm{s}}^2)\overleftarrow{\pa_\eta}\pa_\eta+c_{\rm{s}}^2\pa_\eta^2+2c_{\rm{s}}^2\H \pa_\eta
+\fr{1}{8}(1-3c_{\rm{s}}^2)k^2\right.\nonumber\\
\left.-\fr{1}{2}k'\mu(k-k'\mu)+\fr{3}{4}(1+c_{\rm{s}}^2)k^{'2}
\right\}D(\eta,|\vec{k}-\vec{k}'|),\label{f2}
\end{align}
where $\overleftarrow{\pa_\eta}$ is supposed to differentiate \textit{only} $D(\eta,k')$ in the left.

Introducing the power spectrum of tensor perturbations as
\be
\langle h^r(\vec{k})h^{s*}(\vec{K})\rangle
=\fr{2\pi^2}{k^3}\delta(\vec{k}-\vec{K})\delta_{rs}{\cal P}_h(k)
\ee
and assuming $h^r(\vec{k})$ is Gaussian, 
we can obtain the following expression for the correlation of the source:
\be
\langle S(\eta_1,\vec{k})S(\eta_2,\vec{K})\rangle
=\pi\delta(\vec{k}+\vec{K})\sum_{rs}\int d^3\vec{k}'\fr{{\cal P}_h(k'){\cal P}_h(|\vec{k}-\vec{k}'|)}{k^{'3}|\vec{k}-\vec{k}'|^3}
A_{rs}(\eta_1,\vec{k},\vec{k}')A_{rs}(\eta_2,\vec{k},\vec{k}').
\ee
In this paper, we assume the following \umi{delta-function-type tensor} power spectrum:
\be
{\cal P}_h(k)={\cal A}^2k\delta(k-k_p).\label{Ph}
\ee

\begin{figure}[t]
\begin{center}
\includegraphics[width=12cm,keepaspectratio,clip]{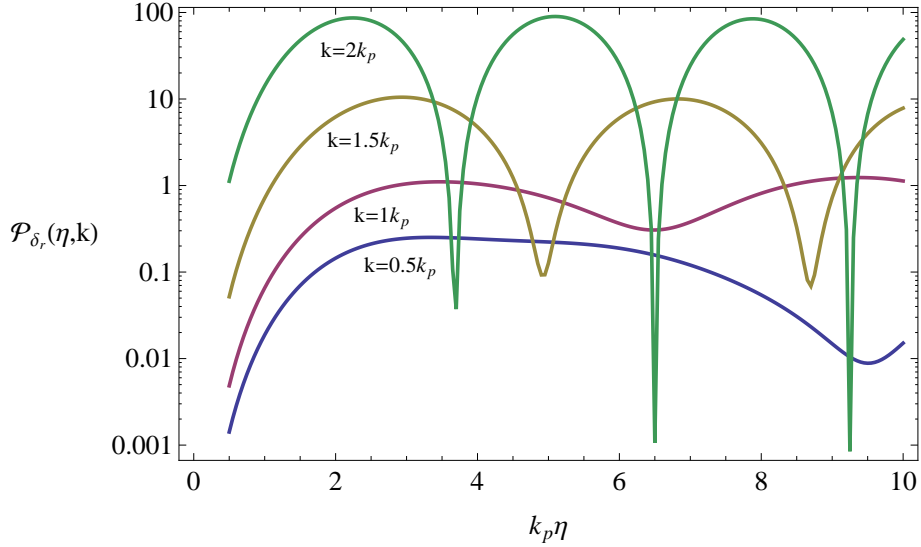}
\end{center}
\caption{The time evolution of the power spectrum for \umi{several modes}, \blue{with $\cal A$ set to unity.}
}
\label{radiation}
\end{figure}

From (\ref{density}) and (\ref{solution}), 
the energy density perturbation can be calculated as
\be
\delta_r\cyan{(\eta, \vec{k})}=
\fr{1+c_{\rm{s}}^2}{c_{\rm{s}}^2\H}\sum_{rs}\int\fr{d^3\vec{k}'}{(2\pi)^{3/2}}h^r(\vec{k}')h^s(\vec{k}-\vec{k}')
F_{rs}(\eta,\vec{k},\vec{k}'),\label{deltar}
\ee
\begin{align}
F_{rs}(\eta,\vec{k},\vec{k}')
&\equiv \int d\tilde{\eta}(\tilde{\eta}/\eta)A_{rs}(\tilde{\eta},\vec{k},\vec{k}')
(\pa_\eta-\H)
g_k(\eta,\tilde{\eta})\nonumber\\
&+D(\eta,k')
\left\{
-\pa_\eta E_1^{rs}+\left(\fr{1}{2}\overleftarrow{\pa_\eta}+\pa_\eta\right)
\left(1-\fr{k'}{k}\mu\right)E_2^{rs}
\right\}
D(\eta,|\vec{k}-\vec{k}'|).\label{Frs}
\end{align}
The power spectrum is defined by
\be
\langle\delta_r(\eta,\vec{k})\delta_r^{*}(\eta,\vec{K})\rangle_{\rm{C}}
\equiv
\langle\delta_r(\eta,\vec{k})\delta_r^{*}(\eta,\vec{K})\rangle-\langle\delta_r(\eta,\vec{k})\rangle
\langle\delta_r^{*}(\eta,\vec{K})\rangle=
\frac{2\pi^2}{k^3}\delta(\vec{k}-\vec{K}){\cal P}_{\delta_r}(\eta,k)\label{power}
\ee
and \umi{is obtained} as follows:
\be
{\cal P}_{\delta_r}(\eta,k)=\left(\fr{1+c_{\rm{s}}^2}{c_{\rm{s}}^2}\right)^2{\cal A}^4\left(\fr{k}{k_p}\right)^2\eta^{2}
\Theta\left(1-\fr{k}{2k_p}\right)
\sum_{rs}F_{rs}\left(\eta,k,k_p,\fr{k}{2k_p}\right)^2.
\ee
The time evolutions of \sora{this} power spectrum for a few modes 
are shown in Fig. \ref{radiation}, where ${\cal A}$ is set to unity. 
The power spectrum takes the maximum value shortly after the horizon crossing of each $k$ mode \blue{($k\eta=1$)}. 
After \blue{reaching} the maximum, it starts \ao{oscillations} with the amplitude almost constant, 
similarly to the behavior in the standard linear cosmological perturbation theory. 
This is because the \souta{first order } tensor perturbations decay after the horizon crossing, 
and so do the source terms, \souta{leading to the reduction of } \sora{and then }our fundamental equations \umi{for scalar perturbations} \sora{are reduced }to 
the standard ones in the linear theory. 
\clearpage
\ao{
\section{Upper bounds on PGWS from PBHs}
}
In order to place upper bounds on tensor modes from PBHs, 
the abundance of PBHs needs to be \umi{related to the primordial tensor power spectrum}, which \souta{is \umi{done}} \sora{can be accomplished} by 
integrating the probability density function (PDF) of the \umi{induced} density perturbation averaged over the horizon. 
In the following we \blue{first} estimate the moment when the PBH formation is most efficient \blue{for each \souta{value of } $k_p$} by calculating the dispersion of the \umi{induced} density perturbation, 
and \blue{then} \souta{calculate} \sora{derive} the PDF at this moment. 

\begin{figure}[t]
\begin{center}
\includegraphics[width=10cm,keepaspectratio,clip]{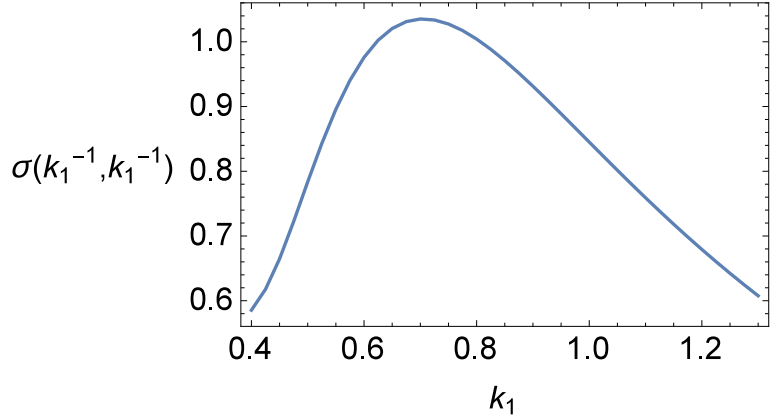}
\end{center}
\caption{The dependence of $\sigma(\eta=k_1^{-1},R=k_1^{-1})$ on $k_1$, 
\umi{with ${\cal A}=1$ and $k_p=1$.}
}
\label{sigma}
\end{figure}

\ao{Let us begin by noting that} the average $\langle\delta_r(\eta,\vec{x})\rangle$ is nonzero, 
since the density perturbation is generated by the \strike{second-order} \souta{effects of } tensor \souta{modes} \sora{perturbations}.
\umi{To evaluate this average} we introduce $f_3$ and $f_4$ by rewriting $F_{rs}$ as
\be
F_{rs}(\eta,\vec{k},\vec{k}')=f_3(\eta,\vec{k},\vec{k}')E_1^{rs}+f_4(\eta,\vec{k},\vec{k}')E_2^{rs}, 
\ee
where the explicit forms of $f_3$ and $f_4$ can be obtained by using (\ref{Ars}), though 
the integration over $\eta$ can not be done analytically for general $\vec{k}$:
\be
f_3(\eta,\vec{k},\vec{k}')=\int d\tilde{\eta}(\tilde{\eta}/\eta)f_1(\tilde{\eta},\vec{k},\vec{k}')
(\pa_\eta-\H)
g_k(\eta,\tilde{\eta})-D(\eta,k')
\pa_\eta 
D(\eta,|\vec{k}-\vec{k}'|),
\ee
\be
f_4(\eta,\vec{k},\vec{k}')=\int d\tilde{\eta}(\tilde{\eta}/\eta)f_2(\tilde{\eta},\vec{k},\vec{k}')
(\pa_\eta-\H)
g_k(\eta,\tilde{\eta})\green{+}
D(\eta,k')\left(\fr{1}{2}\overleftarrow{\pa_\eta}+\pa_\eta\right)
\green{\left(1-\frac{k'}{k}\mu\right)}
D(\eta,|\vec{k}-\vec{k}'|).
\ee
Since only the zero-mode $\delta_r(\eta,\vec{k}=\vec{0})$ contributes to $\langle\delta_r(\eta,\vec{x})\rangle$, 
we need $f_3$ and $f_4$ only in \sora{the limit of} $\vec{k}\rightarrow \vec{0}$\souta{ limit}, which are, under the assumption of the delta-function\umi{-type} power spectrum (\ref{Ph}),
\be
f_3=0,\quad f_4=-\frac{-1+2k_p^2\eta^2+\cos(2k_p\eta)}{24k_p^2\eta^3}.
\ee
Hence, 
\begin{align}
&\langle\delta_r(\eta,\vec{x}=\vec{0})\rangle\nonumber\\
&=\int \frac{d\vec{k}^3}{(2\pi)^{3/2}}\fr{1+c_{\rm{s}}^2}{c_{\rm{s}}^2\H}\int\fr{d^3\vec{k}'}{(2\pi)^{3/2}}\frac{2\pi^2}{k_p^3}\delta(\vec{k}){\cal A}^2k_p\delta(k'-k_p)
f_4(\eta,\vec{k}=\vec{0},\vec{k}')\times \green{(2-(-2))}\nonumber\\
&=-\frac{(1+c_{\rm{s}}^2){\cal A}^2}{6c_{\rm{s}}^2k_p^2\eta^2}\{-1+2k_p^2\eta^2+\cos(2k_p\eta)\}.
\end{align}
When $k_p\eta\gg 1$, the time average of this \ao{quantity} asymptotes to
\be
\langle\delta_r\rangle=-\frac{(1+c_{\rm{s}}^2)}{3c_{\rm{s}}^2}{\cal A}^2,\label{average}
\ee
\green{while $\langle\delta_r\rangle\rightarrow 0$ for $k_p\eta\rightarrow 0$\footnote{\sora{Strictly speaking this effect may be taken into account in the background 
Friedmann equations (\ref{zero1}) and (\ref{zero2}), but ${\cal A}^2$ is mostly less than 0.1 from 
Fig. \ref{constraint}, so the correction to the upper bounds would be $\sim 0.1$ at most, while a rigorous treatment of this effect would greatly complicate analysis. 
Hence we neglect this effect. 
}
}.}
We denote the density perturbation averaged over a sphere with comoving radius $R$ by $\delta_r(\eta,\vec{x},R)$, 
the dispersion of which is related to the power spectrum as follows:
\be
\sigma(\eta,R)\equiv
(\langle\delta_r(\eta,\vec{x},R)^2\rangle-\langle\delta_r(\eta,\vec{x})\rangle^2)^{1/2}
=\left(\int \fr{dk}{k}W^2(kR){\cal P}_{\delta_r}(\eta,k)\right)^{1/2},\label{rootmean}
\ee
where $W$ is the Fourier transform of the top-hat window function: $W(x)=3(\sin x-x\cos x)/x^3$. 
\blue{Figure \ref{sigma} shows that}
the dispersion of the density perturbation at the horizon crossing of some mode $k_1$ 
smoothed over the horizon scale at that moment (namely, $\eta=k_1^{-1}$), $\sigma(\eta=k_1^{-1},R=k_1^{-1})$, is maximum and is $\sim {\cal A}^2$ at around $k_1\sim \green{0.7}k_p$.
\sora{That is, PBHs are formed most efficiently}\souta{\blue{This} means the PBH formation is most efficient} at around this moment, 
\souta{so} \sora{and therefore} we restrict our attention to this moment in the following. 

To determine the abundance of PBHs, the 
PDF of the density perturbations is necessary. 
\umi{Often} the PDF of the density perturbations is assumed to be Gaussian, 
but in our \souta{situation} \sora{problem} it is highly non-Gaussian, 
since \souta{the density perturbation is} \sora{density perturbations are} generated \souta{due to the second-order effects of} \sora{by \strike{second-order}} tensor perturbations, 
whose statistical properties are assumed to be Gaussian. 
We can \sora{in principle determine} \souta{\umi{construct}} the PDF of $\delta_r$ by randomly generating the Fourier modes of GWs $\{h^r(\vec{k})\}$ repeatedly (for the details 
see Appendix \blue{C}), whose result is shown in 
Fig. \ref{pdf}.
\begin{figure}[h]
\begin{center}
\includegraphics[width=13cm,keepaspectratio,clip]{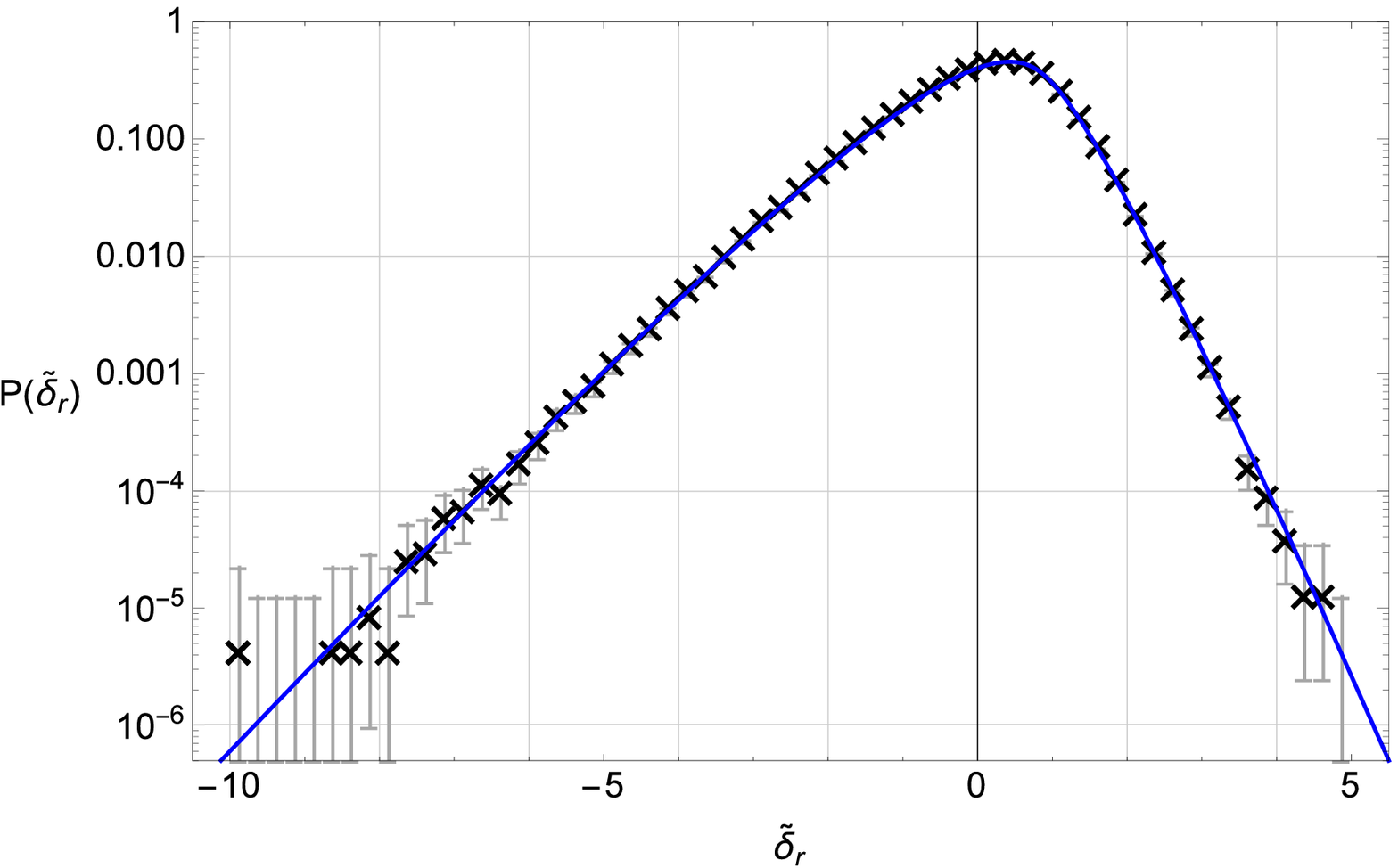}
\end{center}
\caption{
The PDF of \green{$\tilde{\delta}_r\equiv(\delta_r-\langle\delta_r\rangle)/{\cal A}^2$} for \blue{a million} realizations of $\{h^r(\vec{k}_i)\}$ (see Appendix \umi{C} for the details). 
\souta{Here ${\cal A}$ is set to unity. }
The curve is the approximate PDF of \green{$\tilde{\delta}_r$} given by (\ref{last}).
}
\label{pdf}
\end{figure}
The PDF $P(\delta_r)$ of $\delta_r$ \umi{thus obtained} turns out to be well approximated by the formula (\ref{last}). 
\souta{The} \sora{Then the} fraction of the volume which has collapsed into PBHs at \souta{the moment of } \umi{their} formation is \souta{estimated by }
\begin{equation}
\beta=\int_{\delta_{r,\mathrm{th}}}^\infty \tilde{P}(\delta_r-\langle\delta_r\rangle)d\delta_r=\int_{\delta_{r,\mathrm{th}}/{\cal A}^2}^\infty 
P(\tilde{\delta}_r)d\tilde{\delta}_r,
\end{equation}
where $\delta_{r,\mathrm{th}}$ is the threshold\souta{ amplitude required for} \souta{the formation of PBHs} \sora{of PBH formation}, 
in the following assumed to be $\delta_{r,\mathrm{th}}\ao{=} 0.4$ \cite{Polnarev:2006aa,Nakama:2013ica}\footnote{
In these papers the initial conditions of numerical simulations were given in terms of\souta{ the} curvature profile\sora{s} in the limit of the vanishing ratio of 
the Hubble radius to the \souta{scale} \sora{radius} of \souta{perturbation} \sora{perturbed regions}\souta{, which is related to the radial profile of the curvature perturbation}. 
In the present work scalar perturbations are sourced by \strike{second-order} tensor perturbations, and \souta{so} \sora{hence} strictly speaking the \souta{threshold values} \sora{formation conditions} 
obtained \souta{in these numerical simulations} \sora{there} may not be directly applied. A more precise treatment \souta{may} \sora{would} require dedicated numerical simulations, which is 
beyond the scope of this work. \souta{It would be worthwhile to mention here that t} \sora{T}he energy density of \sora{P}GWs is expected to \souta{help} \sora{promote} gravitational collapse, 
in light of previous works on direct collapse of nonlinear \souta{GWs} \sora{gravitational waves}, mentioned in the footnote$\dagger$3 of the Introduction. 
This effect is not taken into account in the present paper, \souta{so} \sora{and therefore} in this sense our upper bounds \ao{would} be conservative.
}. 
This quantity $\beta$ has been constrained \souta{observationally } on various mass\sora{es} \souta{scales} and we use Fig. 9 of \cite{Carr:2009jm}. 
\souta{Upper} \sora{Then upper} bounds \blue{on ${\cal A}^2$} for each\souta{ value of} $\beta$, corresponding to different masses of PBHs, \souta{can be obtained, 
which is} \sora{are} shown in Fig. \ref{constraint}\souta{. 
Here,}\sora{, in which} upper bounds are shown as a function of $k_p$, using the following relation between 
\souta{the mass of PBHs} \sora{the PBH mass} and the comoving \souta{scale} \sora{wave number} of perturbations: 
\be
M_{\rm{PBH}}=2.2\times10^{13}M_\odot\left(\frac{k}{1\rm{Mpc}^{-1}}\right)^{-2}.
\ee
The dependence of the upper bounds on the \souta{perturbation scales} \sora{comoving wave number} is logarithmically weak\souta{, 
which can be understood from} \sora{owing to} the exponential dependence of the PDF on $\delta_r$ and hence on ${\cal A}^2$ for $\delta_r\sora{\simeq \delta_{r,\mathrm{th}}}$. 

\begin{figure}[h]
\begin{center}
\includegraphics[width=13cm,keepaspectratio,clip]{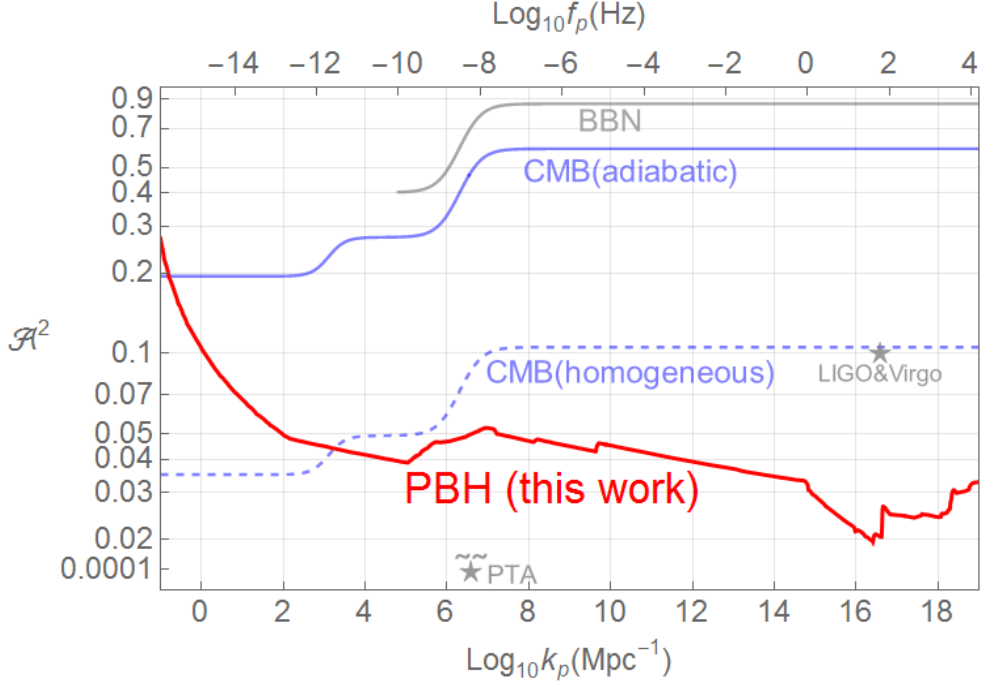}
\end{center}
\caption{
Upper bounds on ${\cal A}^2$ as a function of $k_p$ using PBHs and other methods, \ao{also shown in \cite{Nakama:2015nea}.}
}
\label{constraint}
\end{figure}

Let us compare \souta{this PBH bound} \sora{these PBH bounds} with \umi{other bounds}. 
\umi{We begin by rederiving} the formula for the energy density of gravitational waves $\rho_{\mathrm{GW}}$ on subhorizon scales. 
Noting that $\rho_{\mathrm{GW}}=-\langle S_1\rangle/4\pi Ga^2$ from (\ref{timetime}), where the brackets here imply 
temporal and spatial average (see e.g. \cite{Maggiore} for more details), let us rewrite $\langle S_1/a^2\rangle$ in the following. 
By integration by parts and using (\ref{linear}), 
\begin{align}
\left\langle \frac{S_1}{a^2}\right\rangle
&=
\left\langle
\frac{1}{a^2}\left(
-\frac{1}{4}h_{ij}'h^{ij'}-\frac{3}{2}{\cal H}h_{ij}h^{ij'}+\frac{1}{4}h_{ij}h^{ij''}
\right)
\right\rangle\nonumber\\
&=\left\langle
-\frac{1}{4}\dot{h}_{ij}\dot{h}^{ij}-\frac{5\dot{a}}{4a}h_{ij}\dot{h}^{ij}+\frac{1}{4}h_{ij}\blue{\ddot{h}^{ij}}
\right\rangle
\blue{\simeq}-\frac{1}{2}\langle\dot{h}_{ij}\dot{h}^{ij}\rangle,
\end{align}
hence\footnote{
If we define tensor perturbations without the factor 2 in front of $h_{ij}$ in (\ref{metric}), 
then we arrive at, instead of (\ref{rhogw}), the formula often used in the literature: $\rho_{\rm{GW}}=\langle\dot{h}_{ij}\dot{h}^{ij}\rangle/32\pi G$.
}, 
\be
\rho_{\rm{GW}}=\frac{\langle \dot{h}_{ij}\dot{h}^{ij}\rangle}{8\pi G}.\label{rhogw}
\ee
Assuming the delta-function\umi{-type} power spectrum (\ref{Ph}), 
\be
\rho_{\rm{GW}}=\frac{{\cal A}^2\langle\dot{D}(\eta,k_p)^2\rangle}{2\pi G}
=\frac{{\cal A}^2}{2\pi Ga^2}\left\langle\left(\frac{\cos k\eta}{\eta}-\frac{\sin k\eta}{k\eta^2}\right)^2\right\rangle
\sim \frac{{\cal A}^2}{4\pi Ga^2\eta^2}.
\ee
Defining $\rho_{\rm{crit}}\equiv \rho_{\rm{rad}}+\rho_{\rm{GW}}\simeq 3H^2/8\pi G$, 
the following relation \umi{is obtained, used} shortly:
\begin{equation}
\frac{\rho_{\rm{GW}}}{\rho_{\rm{crit}}}=\frac{2}{3}{\cal A}^2.\label{prep}
\end{equation}
The existence of gravitational waves is often \souta{characterized in terms of } \sora{effectively represented by}
the number of \sora{relativistic fermions'} degrees of freedom\souta{ of relativistic \ao{fermions}}
\sora{as follows}. 
First, the total energy density of radiation without gravitational waves \umi{nor dark radiation} is written as
\be
\rho_{\rm{rad}}(T)=\red{\frac{\pi^2}{30}g_*}T^4, 
\ee
\red{where $g_*$} is 
the effective number of degrees of freedom of relativistic species \blue{and at the epoch of BBN it is} given by \cite{Allen:1996vm,Maggiore:1999vm}
\be
g_*=2+\fr{7}{8}\left\{4+2N_\nu\right\},
\ee
where $N_\nu$ is the effective number of degrees of freedom of neutrinos, $N_\nu=3.046$\footnote{
The slight deviation from $N_\nu=3$ arises from the slight heating of neutrinos 
due to the relic interactions between $e^{\pm}$ and neutrinos at the epoch of $e^{\pm}$ annihilations, 
which took place only shortly after the neutrino decoupling \cite{Mangano:2005cc}. 
}.
This is obtained by counting the degrees of freedom of photons, electrons, positrons, and neutrinos. 
\ao{At \umi{the} photon decoupling}, electrons and positrons should not be included. 
\ao{The presence} of \sora{P}GWs (or possibly of dark radiation) \ao{is} \umi{represented} by $\Delta N_{\rm{eff}}$, as a correction to $N_{\nu}$ above. 
In the following we use $\Delta N_{\rm{GW}}$ as \souta{a} \sora{the} contribution of \sora{P}GWs and \ao{relate it to the primordial tensor power spectrum}. 
When \sora{P}GWs are present, the total energy density becomes (noting (\ref{average}))
\be
\rho_{\rm{tot}}=\rho_{\rm{rad}}(T)(1+\langle\delta_r\rangle)+\rho_{\rm{GW}},
\ee
which can be written, with the redefinition of the temperature $T\rightarrow T(1+\langle\delta_r\rangle/4)$, as
\be
\rho_{\rm{tot}}=\rho_{\rm{rad}}(T)+\rho_{\rm{GW}}.
\ee
After the horizon crossing of \sora{P}GWs, \souta{its energy density \blue{scales} as} $\sora{\rho_{\mathrm{GW}}}\propto a^{-4}$,  
while, denoting \sora{by $g_S(T)$} the effective degrees of freedom of relativistic species in terms of entropy at temperature $T$\souta{ by $g_S(T)$}, 
the photon temperature evolves \souta{following} \sora{according to} $g_S(T)T^3a^3=\:$const. (i.e. constant entropy) and \blue{therefore} $\rho_{\rm{rad}}\propto \red{g_*T^4\sim 1/a^4g_S^{1/3}}$ 
(see e.g. \cite{Maggiore:1999vm}). 
Then, defining $\Omega_{\rm{GW}}\red{\equiv\rho_{\rm{GW}}/\rho_{\rm{crit}}\simeq \rho_{\rm{GW}}/\rho_{\rm{rad}}}$, 
\be
\Omega_{\rm{GW}}(T)=\left(\frac{g_S(T)}{g_S(T_{\rm{in}})}\right)^{\red{1}/3}\Omega_{\rm{GW}}(T_{\rm{in}}),\label{dilution}
\ee
where $T_{\rm{in}}\blue{=T_{\rm{in}}(k_p)}$ is the temperature of radiation \souta{at the moment of the horizon crossing of GWs with comoving \strike{wave number}\sea{wave number} $k_p$} \sora{when PGWs with comoving wave number $k_p$ reenter the horizon}, and 
$T<T_{\rm{in}}$\footnote
{
In \cite{Kuroyanagi:2014nba} the following convenient fitting function is shown: 
\be
g_S(T_{\rm{in}}(k))=g_{S0}\left\{
\frac{A+\tanh\left[-2.5\log_{10}k/2\pi f_1\right]}{A+1}
\right\}
\left\{
\frac{B+\tanh\left[-2.0\log_{10}k/2\pi f_2\right]}{B+1}
\right\},
\ee
where $A=(-1-g_{\rm{BBN}}/g_{S0})/(-1+g_{\rm{BBN}}/g_{S0})$, $B=(-1-g_{\rm{max}}/g_{\rm{BBN}})/(-1+g_{\rm{max}}/g_{\rm{BBN}})$, $g_{S0}=3.91$, $g_{\rm{BBN}}=10.75$,
$f_1=2.5\times 10^{-12}\rm{Hz}$ and $f_2=6.0\times 10^{-9}\rm{Hz}$. 
As for $g_{\rm{max}}$ following \cite{Kuroyanagi:2014nba} we assume the sum of the \souta{standard-model} \sora{Standard Model} particles, $g_{\rm{max}}=106.75$. 
Note that $k/2\pi f_1=k/(1.6\times 10^{-3}\rm{pc}^{-1})$ and $k/2\pi f_2=k/(3.9\rm{pc}^{-1})$.
}.
\ao{At the epoch of BBN,}
the contribution of \sora{P}GWs is \ao{represented} by $\Delta N_{\rm{GW}}$ as follows;
\be
\rho_{\rm{rad}}(T)+\rho_{\rm{GW}}(T)
=\green{\f{\pi^2}{30}}
\left[
2+\frac{7}{8}
\left\{
4+2(N_{\rm{sta}}+\Delta N_{\rm{GW}})
\right\}
\right]T^4,
\ee
which leads to
\be
\rho_{\rm{GW}}(T)=\rho_{\rm{rad}}(T)\times\frac{7}{8}\times 2\times \Delta N_{\rm{GW}}(T)/
\left\{
2+\frac{7}{8}(4+2N_{\rm{sta}})
\right\}\fallingdotseq \frac{7}{43}\rho_{\rm{rad}}\Delta N_{\rm{GW}}(T).
\ee
Since
\be
\Omega_{\rm{GW}}(T_{\rm{in}})\simeq
\fr{2}{3}{\cal A}^2\label{initial}
\ee
\green{from (\ref{prep}), }
$\Delta N_{\rm{GW}}(T)$ can be written as 
\be
\Delta N_{\rm{GW}}(T)=\fr{43}{7}\Omega_{\rm{GW}}(T)=\frac{86}{21}{\cal A}^2
\left(\frac{g_S(T)}{g_S(T_{\rm{in}})}\right)^{\red{1}/3}.\label{DeltaNGW}
\ee
An upper bound on \blue{$\Delta N_{\rm{eff}}$}, $\Delta N_{\rm{eff}}<\Delta N_{\rm{upper}}$, is usually translated into an upper bound on 
$\Delta N_{\rm{GW}}$, $\Delta N_{\rm{GW}}<\Delta N_{\rm{upper}}$. 
\ao{As} is also mentioned in the Introduction, 
\blue{in doing so} \souta{it is assumed} \sora{we assume} that any physical mechanisms, both known and unknown, \souta{contribute positively to} \sora{increase} $N_{\rm{eff}}$, 
\ao{but} at least there are examples where $N_{\rm{eff}}$ decreases \cite{Ichiki:2002eh,Apostolopoulos:2005at,Maartens:2010ar}.
With this in mind, the requirement $\Delta N_{\rm{GW}}<\Delta N_{\rm{upper}}$ is translated into an upper bound on ${\cal A}^2$ from (\ref{DeltaNGW}) as follows:
\red{
\be
{\cal A}^2\lesssim\frac{21}{86}\left(\frac{g_S(T_{\rm{in}})}{g_S(T)}\right)^{1/3}\Delta N_{\rm{upper}},
\ee
}
with $g_S(T)=g_{\mathrm{BBN}}=10.75.$  

\green{On the other hand, at the \umi{photon} decoupling,
\begin{align}
\rho_{\rm{rad}}(T)+\rho_{\rm{GW}}(T)
=\frac{\pi^2}{30}\nami{2+2\times \frac{7}{8}\maru{\frac{4}{11}}^{4/3}(N_\nu+\Delta N_{\rm{GW}})},
\end{align}
which yields
\begin{equation}
\Omega_{\rm{GW}}(T)=\frac{2\times\f{7}{8}\maru{\frac{4}{11}}^{4/3}}{2+2\times\f{7}{8}\maru{\frac{4}{11}}^{4/3}N_\nu}\Delta N_{\rm{GW}}
\simeq 0.13\Delta N_{\rm{GW}}.
\end{equation}
So in this case we find
\be
{\cal A}^2<0.13\times\frac{3}{2}\times\Delta N_{\rm{upper}}\left(\frac{g_S(T_{\rm{in}})}{g_S(T)}\right)^{1/3},
\ee
\blue{with $g_S(T)=g_{S0}=3.91.$}
\umi{These constraints} depend on \blue{$g_S(T_{\rm{in}})$}, 
which one may \red{regard as} a drawback of these methods 
since \blue{it} is uncertain especially at high temperatures. 
\sora{It is also potentially affected by some entropy production mechanisms}
\souta{It also depends on other potential entropy productions} \cite{Kuroyanagi:2014nba}. 
On the other hand, the PBH constrain\sora{s}\souta{t does} \sora{do} not depend on $g_S$ nor other entropy productions \blue{much}. 
}

In order not to spoil \umi{the} \sora{successful standard} BBN, we follow \cite{Kuroyanagi:2014nba} \umi{and set} 
$\Delta N_{\rm{upper}}=1.65$
as a 95\% C.L. upper limit, 
which is applicable for the scales smaller than the comoving horizon at \souta{the time of \umi{the} }BBN, namely, $6.5\times 10^4\rm{Mpc}^{-1}\lesssim \textit{k}$. 
As for CMB constraints, 
in \cite{Smith:2006nka} the use of homogeneous initial conditions of \sora{P}GWs\sora{' energy density} is advocated for 
\sora{those} generated, for instance, by quantum fluctuations during inflation. 
In this case \sora{we use the} 95 \% upper limit\souta{s are} \sora{of} $\Delta N_{\rm{upper}}=0.18$ \sora{from} \cite{Sendra:2012wh}\footnote
{
One would \souta{get} \sora{obtain} somewhat tighter constraints than those in \cite{Sendra:2012wh} for 
homogeneous initial conditions of \sora{PGWs'} energy density, 
by repeating the analysis of \cite{Sendra:2012wh} using more recent data.
}.
For adiabatic initial conditions of \sora{P}GWs we refer to
\begin{equation}
N_{\rm{eff}}=3.52_{-0.45}^{+0.48}\quad(95\%;\;\;Planck+\rm{WP}+\rm{highL}+H_0+\rm{BAO})
\end{equation}
of \cite{Ade:2013zuv} to set $\Delta N_{\rm{upper}}=1.00$ \cite{Kikuta:2014eja}.

\green{
The current energy density of \sora{PGWs}, $\Omega_{\rm{GW},0}$, is \red{also constrained by LIGO and Virgo, most severely 
\green{in the band $41.5-169.25\rm{Hz}$} as $\Omega_{\rm{GW},0}\lesssim 5.6\times 10^{-6}\times\log(169.25/41.5)\simeq 8\times 10^{-6}$ \cite{Aasi:2014zwg}}. 
\blue{Since} $\Omega_{\rm{GW},0}\sim (4/100)^{\red{1}/3}2{\cal A}^2/3z_{\rm{eq}}\sim 7.6\times 10^{-5}{\cal A}^2$ 
($z_{\rm{eq}}\sim 3000$ is the redshift at the matter-radiation equality, and \red{the factor $z_{\rm{eq}}^{-1}$} reflects $\Omega_{\rm{GW}}\propto (1+z)/(1+z_{\rm{eq}})$ during \red{\umi{the} matter-dominated era}), 
we \blue{obtain} ${\cal A}^2\lesssim \green{0.1}$\footnote{
  \cyan{
  \blue{Though not included in our analysis,} they also obtained weaker upper bounds on a \umi{few} frequency ranges other than the one around $\sim 100$Hz.
  \umi{Also, strictly} speaking in \cite{Aasi:2014zwg} some power-low \souta{frequency dependence \umi{of GWs}} \sora{spectrum of gravitational waves} is assumed in each band, 
  and so their results may not be directly translated into constraints on a narrow peak in the power spectrum we consider. 
  Indeed in \cite{Nishizawa:2015oma} an optimal analysis method is discussed to search for a sharp emission line of \sora{gravitational waves}, which 
  can increase the signal-to-noise ratio by up to a factor of seven.
  Namely, our comparison here \blue{may be crude}, but it is sufficient for our purposes. 
  \green{The same applies \green{to} the comparison with PTA. }
  }  
}.
}

\green{
Pulsar timing arrays (PTAs) have also been used to constrain \sora{P}GWs. Following \cite{Kuroyanagi:2014nba} 
we use the most stringent upper bound around $f=5.72\green{\times 10^{-9}}$Hz ($\sim \green{4\times 10^6}\rm{Mpc}^{-1}$), 
$\Omega_{\rm{GW},0}\sim (4/11)^{1/3}2{\cal A}^2/3z_{\rm{eq}}\lesssim \blue{2}\times10^{-8}$, 
which leads to ${\cal A}^2\lesssim \blue{1.3}\times 10^{-4}$.
}

\green{\souta{\ao{GW}}\sora{Ground-based} detectors \green{or PTA experiments} \ao{constrain} \sora{P}GWs on a relatively limited frequency range, while cosmological methods \souta{like} \sora{such as} PBHs probe \sora{P}GWs on a wide range of frequencies, 
and this is another advantage \souta{of PBHs in constraining primordial \sora{P}GWs} \sora{of our new limits} \green{(see Fig. \ref{constraint})}. 
}

These upper bounds as a function of $k_p$ \blue{are} shown in Fig. \ref{constraint} along with the \ao{upper bound from PBHs}.
One may not regard some of \souta{these} \sora{weak} constraints \sora{there} as meaningful, because \souta{upper limits} \sora{they} correspond to 
\souta{the amplitude of GWs which is} (almost) nonlinear \sora{tensor perturbations}. 

\begin{figure}[t]
\begin{center}
\includegraphics[width=13cm,keepaspectratio,clip]{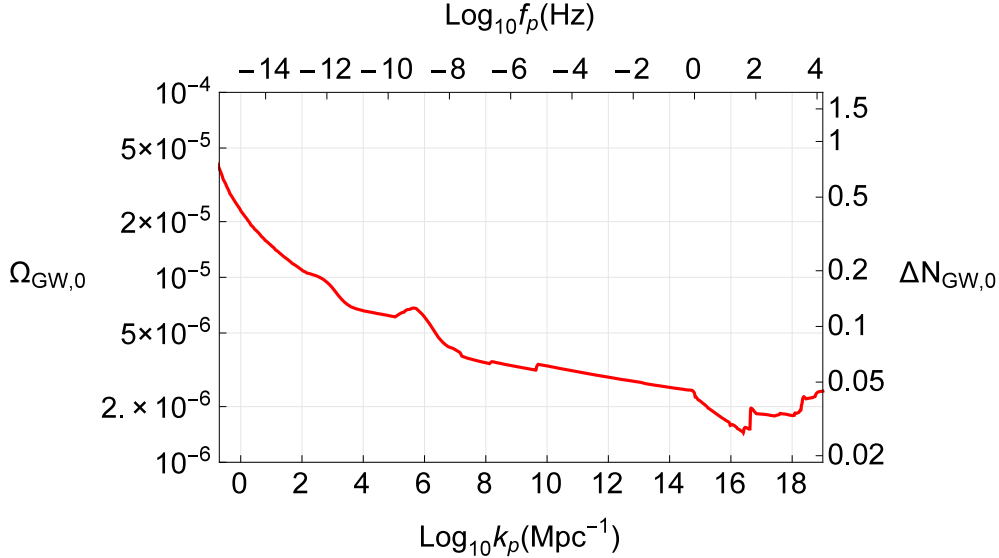}
\end{center}
\caption{
\green{\ao{Upper bounds from PBHs} on $\Omega_{\rm{GW},0}$ \umi{or equivalently} $\Delta N_{\rm{GW},0}$ as a function of $k_p$. }
}
\label{constraint2-2}
\end{figure}

\umi{As shown in Fig. \ref{constraint2-2},}
\ao{the upper bounds from PBHs} can also be expressed in terms of $\Delta N_{\rm{GW}}$ using (\ref{DeltaNGW}), 
and also in terms of the current energy density parameter of \sora{P}GWs, $\Omega_{\rm{GW},0}$, using
\begin{equation}
\Omega_{\mathrm{GW},0}=\frac{2{\cal A}^2}{3z_{\mathrm{eq}}}\left(\frac{g_{S0}}{g_S(T_{\mathrm{in}})}\right)^{1/3}, 
\end{equation}
which follows from (\ref{dilution}) and (\ref{initial}). 
\ao{\souta{If}\sora{Note that if} future experiments reveal the presence of} $\Delta N_{\rm{eff}}$, then \souta{primordial GWs} \sora{PGWs} provide a possible explanation, as well as 
dark radiation. However, if the value of $\Delta N_{\rm{eff}}$ is large, say 0.5, exceeding the \souta{values indicated} \sora{limits shown} in Fig. \ref{constraint2-2}, 
then we may exclude \souta{primordial GWs} \sora{PGWs} as a candidate thanks to our 
PBH bounds\souta{ on them}\footnote{
There may be a loophole, however. Logically, if \souta{primordial GWs} \sora{PGWs} follow a tremendously non-Gaussian PDF, it \souta{maybe} \sora{may be} possible to realize large $\Delta N_{\mathrm{GW}}$ 
without \souta{producing too may} \sora{overproducing} PBHs.
}. This shows an example of how our \souta{\ao{upper bounds from PBHs}} \sora{new limits} can provide useful \umi{cosmological} information.
\vspace{1cm}
\section{Conclusion}
A novel method \blue{using PBH formation} to probe \ao{primordial} gravitational waves is discussed. 
If the amplitude of tensor perturbations initially on superhorizon scales is very large, 
\sora{substantial }scalar perturbations \sora{are }generated \souta{due to the second-order effects of tensor modes \umi{become important}} \sora{from \strike{second-order} tensor perturbations}. 
If \souta{the amplitude of resultant} \sora{these} induced scalar perturbations \souta{is} \sora{are} too large, 
PBHs are \sora{overproduced, exceeding existing upper limits on their abundance}\souta{produced too abundantly to a level that is inconsistent with various types of existing observations which have 
placed upper bounds on the abundance of PBHs on each mass scale}. 

To constrain tensor modes \ao{by PBHs formed by \souta{the } gravitational collapse} of \souta{the radiation density perturbations} \sora{radiation overdensities}, we have calculated the PDF of the radiation density perturbations, which 
is \umi{in general} highly non-Gaussian since they are sourced by \souta{the terms in the Einstein equations which are second-order in tensor modes $h_{ij}$} \strike{second-order} tensor perturbations.
\umi{Assuming primordial tensor perturbations are Gaussian,}
\ao{an approximate analytic \umi{formula} of the PDF was derived, \umi{which} coincides well with the PDF obtained by a \souta{direct } Monte Carlo simulation.}

Using this PDF we have \sora{constrained}\souta{placed upper bounds on the initial amplitude of tensor modes assuming } a delta-function\umi{-type} power spectrum \sora{of primordial tensor perturbations}. 
Our findings \souta{\umi{can be}} \sora{are} summarized in Fig. \ref{constraint}\sora{.}\souta{, which shows comparison of PBH constraints with other constraints. 
It would be worthwhile \ao{to apply our formalism presented in this paper to} other types of tensor power spectrum such as a blue spectrum. }

PBH constraints are applicable from comoving scales of \souta{$\sim \rm{Gpc}$} \sora{$\sim \rm{Mpc}$} all the way down to those of $\sim \ao{0.1}$m 
if we assume the number of $e$-folds during inflation is sixty. 
\ao{The} exclusion of an overproduction of smallest PBHs ($M_{\rm{PBH}}\lesssim 10^5$\ao{g}) depends on the assumption that 
stable Planck mass relics are left over at the end of Hawking evaporation, which \souta{contribute to} \sora{behave as} cold dark matter (see \cite{MacGibbon:1987my}, \cite{Carr:2009jm} and references therein). 
The range of comoving scales corresponding to $M_{\rm{PBH}}\lesssim 10^5$\ao{g} is roughly $\lesssim 50$ \ao{m}, 
\souta{and so} \sora{namely,} \sora{the} \ao{upper bounds from PBHs} in this range \souta{depend} \sora{are based} on this assumption. 
If Planck mass relics are not \souta{formed} \sora{left over}, 
to what extent an overproduction of PBHs \souta{with $M_{\rm{PBH}}\lesssim 10^5$\ao{g}} \sora{lighter than $10^5$g} is cosmologically problematic is uncertain. 
Such an overproduction of smallest PBHs may lead to an early matter-dominated era, 
during which  PBH binaries are formed and emit \souta{GWs} \sora{gravitational waves}, 
or larger PBHs may form due to merger taking place after the collapse of perturbations of PBHs' density, 
thereby leaving observable traces \cite{Dolgov:2011cq}. 
Therefore, in principle one may still exclude such an \ao{overproduction} of smallest PBHs even without the left over of Planck mass relics 
to fully validate our upper bounds on smallest scales, 
though we do not discuss it in detail here. 

We have used \umi{a} perturbative expansion \umi{based on} small perturbations 
\umi{and therefore} one may be worried about the validity of the PDF, shown in Fig. \ref{pdf}, close to the threshold 
of $\delta_{\rm{th}}\simeq 0.4$ we adopted, \ao{since this value} indicates \sora{that} further nonlinearities may affect. Very naively, next-order corrections would appear in the 
fundamental equations \souta{whose magnitude is} \sora{which are suppressed by} $\sim {\cal O}(h_{ij})\sim \sqrt{0.4}\sim 0.6$, \souta{so} \sora{and this implies that} the upper bounds can be affected by $\sim 60\%$. 
Certainly this estimation is very naive and \souta{so }a more careful estimation would be merited. 
If additional nonlinearities 
\souta{enlarge the amplitude of} \sora{enhance} induced \sora{scalar} perturbations, then our upper bounds\souta{ based on the present formulation} would be conservative. 
To see how \souta{these } \sora{further} nonlinearities\souta{ can} affect \souta{eventual results} \sora{our limits}, one may write down the next-order correction terms, and \sora{then} the behaviors of these terms 
\souta{may} \sora{would} provide insight\souta{ into how additional nonlinearities affect our results}. \umi{A gradient} expansion approach may also be helpful (see e.g. \cite{Harada:2015yda} and references therein), \umi{which} is another perturbative \umi{scheme} based on the smallness 
of the ratio of spatial derivatives to time derivatives for perturbations on superhorizon scales. 
\souta{This} \sora{It} is valid only on superhorizon scales, but \souta{this can treat } nonlinear perturbations \sora{can be treated}, relevant to PBH formation.
If one compares the amplitude of induced perturbations obtained by \umi{a} gradient expansion approach and 
that we have obtained, one would gain insight into how nonlinearities might affect. 
However, \souta{the gradient expansion approach itself} \sora{this approach} is not perfect either, since it does not allow us to evolve perturbations 
up to the moment of \souta{the horizon crossing of perturbations under consideration} \sora{their horizon reenty}, \souta{which is required} \sora{necessary} to calculate the 
probability of PBH formation. Refining our results further would be a formidable task\souta{, which is beyond the scope of this work}. 
The present formulation would be acceptable, providing moderately precise and potentially conservative bounds, 
for our purpose here to propose a novel method to constrain primordial tensor perturbations on small scales from PBHs 
with detailed calculations for the first time. 
\sora{Let us emphasize that, t}\souta{T}hough \ao{upper bounds on scalar perturbations from PBHs} have long been known, probably since \cite{Carr:1993aq}, 
we have newly found \blue{upper bounds from PBHs} on \textit{tensor} perturbations as well.

We have also assumed Gaussianity of primordial tensor perturbations, 
but PBH constraints on tensor perturbations \sora{naturally} depend on \souta{statistics} \sora{their statistical properties}\souta{ of tensor perturbations}, determining \souta{the statistics} \sora{those} of induced density perturbation\sora{s}, 
just as PBH constraints on \textit{scalar} perturbations depend on the \souta{statistics} \sora{statistical properties} of scalar perturbations \cite{Byrnes:2012yx}. 
If high-$\sigma$ realizations of tensor perturbations are suppressed (enhanced) in comparison to a Gaussian case, PBH constraints on tensor perturbations are tighter (weaker). 

We have restricted attention to PBH formation as a result of direct collapse of radiation density perturbations induced \souta{\ao{at second order} in} \sora{by \strike{second-order}} tensor perturbations, 
but \souta{these density perturbations} \sora{they} would also dissipate to induce CMB spectral distortion, and hence constraints on 
CMB spectral distortion \souta{would} \sora{can} also be used to probe tensor perturbations. 
Furthermore, \strike{second-order} tensor perturbations naturally induce perturbations in \umi{the dark matter energy density} as well, and if \souta{the amplitude of 
\umi{them} is} \sora{they are} \umi{sufficiently} large, \souta{it leads to} \sora{they result in a} substantial formation of \ao{what are sometimes called ultracompact minihalos}, \umi{\souta{mihihalos}\sora{small dark matter halos} formed \souta{at redshifts } well before the standard structure formation, say $z\sim 1000$.} That is, (potential) constraints on \ao{them} can also be 
translated into upper bounds on tensor perturbations, which will be explored elsewhere \cite{nsprep}. 

\sora{
Lastly, our analysis based on the delta-function spectrum also has implications on \strike{constraints on} \sea{constraining} other types of tensor power spectra. To see this let us consider the following blue spectrum:
\begin{equation}
{\cal P}_h(k)=r{\cal P}_\zeta(k_{\mathrm{ref}})\left(\frac{k}{k_{\mathrm{ref}}}\right)^{n_T}\:\:\: \mathrm{for}\:\:\: k<k_{\mathrm{max}},
\end{equation}
where ${\cal P}_\zeta$ is the dimensionless power spectrum of the curvature perturbation, $r$ is the tensor-to-scalar ratio, $k_{\mathrm{ref}}$ is some reference \strike{wave number} \sea{wave number} and $n_T\sea{>0}$ is the tensor spectral index. 
\sea{If $n_T$ is relatively large, say, the upper limit $0.45$ obtained below, 
the amplitude of gravitational waves is mostly determined by the modes with wave number close to $k_{\mathrm{max}},$  and as a result the above spectrum can roughly be regarded as equivalent to a delta-function spectrum (\ref{Ph}) with $k_p=k_{\mathrm{max}}$ and
}
\strike{As an illustration, in the following we take $r=0.01, k_{\mathrm{ref}}=0.01\mathrm{Mpc}^{-1},{\cal P}(k_{\mathrm{ref}})=2.2\times 10^{-9}$ and $k_{\mathrm{max}}=10^{18}\mathrm{Mpc}^{-1}$.} \strike{Our constraints on the delta-function spectrum even provide a rough estimation on an upper limit on $n_T$ as follows. The above spectrum can be roughly equivalent to the presence of delta-function power with amplitude }
\begin{equation}
{\cal A}^2=
\int_{\mathrm{e}^{-1}k_{\mathrm{max}}}^{k_{\mathrm{max}}}r{\cal P}_\zeta(k_{\mathrm{ref}})\left(\frac{k}{k_{\mathrm{ref}}}\right)^{n_T}\frac{dk}{k}\sea{.}\label{integral}
\end{equation}
\sea{As an illustration, in the following we take $r=0.01, k_{\mathrm{ref}}=0.01\mathrm{Mpc}^{-1},{\cal P}(k_{\mathrm{ref}})=2.2\times 10^{-9}$ and $k_{\mathrm{max}}=10^{18}\mathrm{Mpc}^{-1}$.}
Using ${\cal A}^2\lesssim 0.02$ at around $10^{18} \mathrm{Mpc}^{-1}$ from Fig. \ref{constraint}, we obtain $n_T\lesssim 0.45$ from the above\footnote{
The limit on $n_T$ is relatively insensitive to the lower bound $k_{\mathrm{min}}$ of the integration of (\ref{integral}): we obtain $n_T\lesssim 0.47$ for $k_{\mathrm{min}}=2k_{\mathrm{max}}/3$, 
and $n_T\lesssim 0.43$ for $k_{\mathrm{min}}=0$.
}. Here we have neglected the modes with $k<\mathrm{e}^{-1}k_{\mathrm{max}}$, but this probably makes this limit on $n_T$ conservative, since contributions of those modes also create density perturbations collapsing to PBHs. Having said that, dedicated calculations for other types of tensor power spectra would be merited. 
}
\color{black}
\section*{Appendix A: \\Derivation of evolution equations for \sora{induced} scalar perturbations
\souta{\umi{induced by} second-order tensor perturbations}}
In this appendix we derive \souta{our} \sora{the} fundamental equations for scalar perturbations induced by \strike{second-order} tensor perturbations. 
\umi{First we derive the parts of the equations involving only scalar perturbations, and then we derive the source terms, second order in tensor perturbations.}
We have also checked the expressions below by a \sora{Mathematica} package\souta{ for Mathematica}, xPand \cite{Pitrou:2013hga}.
\subsection*{Scalar perturbation}
We use the formulation of \cite{Weinberg}, in which the metric \ao{is} decomposed as
\begin{equation}
g_{\mu\nu}=\bar{g}_{\mu\nu}+\ao{\delta g}_{\mu\nu},
\end{equation}
\begin{equation}
\bar{g}_{00}=-1,\quad \bar{g}_{i0}=\bar{g}_{0i}=0,\quad \bar{g}_{ij}=a^2\delta_{ij}.
\end{equation}
The components of the perturbed Ricci tensor are expressed in terms of $\ao{\delta g}_{\mu\nu}$
as follows \cite{Weinberg}\footnote{
\label{sign}In \cite{Weinberg} the Ricci tensor is defined by 
\begin{equation}
R_{\mu\nu}\equiv \Gamma^\lambda_{\mu\lambda,\nu}
-\Gamma^\lambda_{\mu\nu,\lambda}
+\Gamma^\kappa_{\mu\lambda}\Gamma^\lambda_{\nu\kappa}
-\Gamma^\kappa_{\mu\nu}\Gamma^\lambda_{\lambda\kappa}.\label{Ricci}
\end{equation}
With this definition, the Einstein equations are written as
\begin{equation}
R_{\mu\nu}-\frac{1}{2}g_{\mu\nu}R=-8\pi GT_{\mu\nu}.
\end{equation}
If we adopt another definition of the Ricci tensor, which is minus that 
of (\ref{Ricci}), then the sign of the right hand side of the \sora{above} Einstein equations 
should be flipped. 
We adopt the former definition in this section following \cite{Weinberg}, 
but in the next section we adopt the latter definition. 
}:
\begin{align}
\delta R_{jk}=&-\frac{1}{2}\ao{\delta g}_{00,jk}-(2\dot{a}^2+a\ddot{a})\ao{\delta g}_{00}\delta_{jk}-\frac{1}{2}
a\dot{a}\dot{\ao{\delta g}}_{00}\delta_{jk}\nonumber\\
&+\frac{1}{2a^2}(\Delta \ao{\delta g}_{jk}-\ao{\delta g}_{ik,ij}-\ao{\delta g}_{ij,ik}+\ao{\delta g}_{ii,jk})\nonumber\\
&-\frac{1}{2}\ddot{\ao{\delta g}}_{jk}+\frac{\dot{a}}{2a}(\dot{\ao{\delta g}}_{jk}-\dot{\ao{\delta g}}_{ii}\delta_{jk})
+\frac{\dot{a}^2}{a^2}(-2\ao{\delta g}_{jk}+\ao{\delta g}_{ii}\delta_{jk})+\frac{\dot{a}}{a}\ao{\delta g}_{i0,i}\delta_{jk}
\nonumber\\
&+\frac{1}{2}(\dot{\ao{\delta g}}_{k0,j}+\dot{\ao{\delta g}}_{j0,k})+\frac{\dot{a}}{2a}(\ao{\delta g}_{k0,j}+\ao{\delta g}_{j0,k}),
\end{align}
\begin{align}
\delta R_{0j}=\delta R_{j0}=&\frac{\dot{a}}{a}\ao{\delta g}_{00,j}+\frac{1}{2a^2}(\Delta \ao{\delta g}_{j0}-\ao{\delta g}_{i0,ji})
-\left(\frac{\ddot{a}}{a}+\frac{2\dot{a}^2}{a^2}\right)\ao{\delta g}_{j0}\nonumber\\
&+\frac{1}{2}\frac{\partial}{\partial t}\left[\frac{1}{a^2}(\ao{\delta g}_{kk,j}-\ao{\delta g}_{kj,k})\right],
\end{align}
\begin{align}
\delta R_{00}=&\frac{1}{2a^2}\Delta \ao{\delta g}_{00}+\frac{3\dot{a}}{2a}\dot{\ao{\delta g}}_{00}-\frac{1}{a^2}
\dot{\ao{\delta g}}_{i0,i}\nonumber\\
&+\frac{1}{2a^2}\left[\ddot{\ao{\delta g}}_{ii}-\frac{2\dot{a}}{a}\dot{\ao{\delta g}}_{ii}+2\left(\frac{\dot{a}^2}{a^2}-\frac{\ddot{a}}{a}\right)\ao{\delta g}_{ii}\right].
\end{align}
The components of the Ricci tensor with mixed indices are expressed in terms of those with doubly covariant indices as follows:
\begin{equation}
\delta R^0_0=-3\frac{\ddot{a}}{a}\ao{\delta g}_{00}-\delta R_{00},
\end{equation}
\begin{equation}
\delta R^0_{i}=-\delta R_{0i}-a^{-2}(2\dot{a}^2+a\ddot{a})\ao{\delta g}_{i0},
\end{equation}
\begin{equation}
\delta R^i_j=a^{-2}\left(2H^2+\frac{\ddot{a}}{a}\right)\ao{\delta g}_{ij}+\frac{1}{a^2}\delta R_{ij}.
\end{equation}
Using these, the Ricci scalar can be calculated as
\begin{align}
a^2\delta R=&-3a\dot{a}\dot{\ao{\delta g}}_{00}-6(\dot{a}^2+a\ddot{a})\ao{\delta g}_{00}-\Delta \ao{\delta g}_{00}
+2\dot{\ao{\delta g}}_{i0,i}+4H\ao{\delta g}_{i0,i}\nonumber\\
&-\ddot{\ao{\delta g}}_{ij}+\frac{2}{3a^2}\Delta \ao{\delta g}_{ii}+2\left(H^2+\frac{\ddot{a}}{a}\right)\ao{\delta g}_{ii}.
\end{align}
In our notation of (\ref{metric}),
\begin{equation}
\ao{\delta g}_{00}=-2\Phi\;,\quad \ao{\delta g}_{i0}=a B_{,i}\;,\quad \ao{\delta g}_{ii}=-6a^2\Psi\;.
\end{equation}
The time-time component of the Einstein \souta{tensor becomes} \sora{equations is}
\begin{equation}
\frac{a^2}{2}G^0_0=\Delta \Psi-3{\cal H}(\Psi'+{\cal H}^2\Phi)-{\cal H}\Delta B=\frac{a^2}{2}\ao{8\pi G\delta \rho},\label{00scalar}
\end{equation}
which recovers the parts of (\ref{timetime}) involving scalar perturbations. 
The time-space component is
\begin{equation}
G^0_i=R^0_i=-\delta R_{0i}-a^{-2}(2\dot{a}^2+a\ddot{a})\ao{\delta g}_{i0}=2\dot{\Psi}_{,i}+2H\Phi_{,i}.
\label{i0scalar}
\end{equation}
So $aG^0_i/2=0$ partially recovers (\ref{timespace}).
The space-space components are
\begin{equation}
\delta G^i_j=a^{-2}\left(2H^2+\frac{\ddot{a}}{a}\right)\ao{\delta g}_{ij}+\frac{1}{a^2}\delta R_{ij}
-\frac{1}{2}\delta R\delta_{ij},
\end{equation}
and this is written in the form $\delta G^i_j=G_1\delta_{ij}+G_{2,ij}$, where 
\begin{align}
-\frac{a^2}{2}G_1=\Psi''+{\cal H}(2\Psi+\Phi)'+(2{\cal H}'+{\cal H}^2)\Phi+
\frac{1}{2}\Delta(\Phi-\Psi+B'+2{\cal H}B),\label{ijscalar1}
\end{align}
\begin{equation}
a^2G_2=\Phi-\Psi+B'+2{\cal H}B.\label{ijscalar2}
\end{equation}
Then, $-a^2G_1/2=a^2\ao{8\pi G\delta p}/2$ partially recovers (\ref{ein3}), 
and $a^2G_{2,ij}=0$ partially recovers (\ref{ein4}).
Also, (\ref{cons1}) and (\ref{cons2}) without the source term can be derived from (5.1.49) and (5.1.48) of \cite{Weinberg}\footnote{
One can also confirm \umi{that}, dropping the source terms originating from \strike{second-order} tensor perturbations, 
Eqs. (\ref{timetime})-(\ref{ein4}), (\ref{cons1}) and (\ref{cons2}) reduce to Eqs. (A.98)-(A.103) of \cite{Baumann:2009ds}.
}.

\subsection*{Tensor perturbation}
Let us consider the following metric
\begin{equation}
ds^2=a^2(\eta)[-d\eta^2+a^2(\delta_{ij}+\ao{\tilde{h}}_{ij})dx^idx^j],\label{metric2}
\end{equation}
\ao{where $\tilde{h}_{ij}$ is two times $h_{ij}$ in (\ref{metric}) and is introduced here for simplicity.}
We decompose the metric (and other tensors below) as $g_{ij}=\bar{g}_{ij}+\delta g_{ij}+\delta^2g_{ij}$, with $\bar{g}_{ij}=a^2\delta_{ij}, \delta g_{ij}=a^2\ao{\tilde{h}}_{ij}, \delta^2 g_{ij}=0$. 
Then, $\bar{g}^{ij}=a^{-2}\delta_{ij}, \delta g_{ij}=-a^2\ao{\tilde{h}}_{ij}, \delta^2 g^{ij}=\ao{\tilde{h}}^{ik}\ao{\tilde{h}}^j_k$. The indices of $\ao{\tilde{h}}_{ij}$ are raised and lowered by $\delta_{ij}$. 
The nonvanishing components of the Christoffel symbol are 
\begin{equation}
\bar{\Gamma}^0_{00}={\cal H},\quad
\bar{\Gamma}^0_{ij}={\cal H}\delta_{ij},\quad
\delta \Gamma^0_{ij}=\frac{1}{2}(\ao{\tilde{h}}_{ij}'+2{\cal H}\ao{\tilde{h}}_{ij}),
\end{equation}
\begin{equation}
\bar{\Gamma}^i_{j0}={\cal H}\delta_{ij},\quad
\delta \Gamma^i_{j0}=\frac{1}{2}\ao{\tilde{h}}^{i'}_j,\quad
\delta^2 \Gamma^i_{j0}=-\frac{1}{2}\ao{\tilde{h}}^{ik}\ao{\tilde{h}}_{kj}',
\end{equation}
\begin{equation}
\delta \Gamma^i_{jk}=\frac{1}{2}(\ao{\tilde{h}}_{ij,k}+\ao{\tilde{h}}_{ik,j}-\ao{\tilde{h}}_{jk,i}),\quad
\delta^2\Gamma^i_{jk}=\frac{1}{2}\ao{\tilde{h}}^{il}(\ao{\tilde{h}}_{lj,k}+\ao{\tilde{h}}_{lk,j}-\ao{\tilde{h}}_{jk,l}).
\end{equation}
The components of the Ricci tensor are
\begin{equation}
\delta^2R_{00}=\frac{1}{2}\ao{\tilde{h}}^{ij}\ao{\tilde{h}}_{ij}''+\frac{1}{4}\ao{\tilde{h}}^{ij'}\ao{\tilde{h}}_{ij}'+\frac{1}{2}{\cal H}\ao{\tilde{h}}^{ij}\ao{\tilde{h}}_{ij}',
\end{equation}
\begin{equation}
\delta^2 R_{i0}=\frac{1}{4}\ao{\tilde{h}}^{jk'}\ao{\tilde{h}}_{jk,i}+\frac{1}{2}\ao{\tilde{h}}^{jk}\ao{\tilde{h}}_{jk,i}'-\frac{1}{2}\ao{\tilde{h}}^{jk}\ao{\tilde{h}}_{ij,k}',
\end{equation}
\begin{equation}
\bar{R}_{ij}=({\cal H}'+2{\cal H}^2)\delta_{ij},
\end{equation}
\begin{equation}
\delta R_{ij}=\frac{1}{2}\ao{\tilde{h}}_{ij}''+{\cal H}\ao{\tilde{h}}_{ij}'+({\cal H}'+2{\cal H}^2)\ao{\tilde{h}}_{ij}-\frac{1}{2}\Delta \ao{\tilde{h}}_{ij},
\end{equation}
\begin{align}
\delta^2R_{ij}=-\frac{{\cal H}}{2}\ao{\tilde{h}}^{kl}\ao{\tilde{h}}_{kl}'\delta_{ij}-\frac{1}{2}\ao{\tilde{h}}_i^{k'}\ao{\tilde{h}}_{kj}'&+\frac{1}{2}\ao{\tilde{h}}^{kl}(\ao{\tilde{h}}_{ij,kl}-\ao{\tilde{h}}_{ik,jl}-\ao{\tilde{h}}_{jk,il})
+\frac{1}{2}\ao{\tilde{h}}^{kl}\ao{\tilde{h}}_{kl,ij}\nonumber\\
&+\frac{1}{4}\ao{\tilde{h}}^{kl}_{\;\;,i}\ao{\tilde{h}}_{kl,j}+\frac{1}{2}\ao{\tilde{h}}_{i}^{k,l}\ao{\tilde{h}}_{jk,l}-\frac{1}{2}\ao{\tilde{h}}_{i}^{k,l}\ao{\tilde{h}}_{jl,k}.
\end{align}
The components of the Ricci tensor with mixed indices are given by 
\begin{equation}
\delta^2R^0_0=-a^{-2}\delta^2R_{00},\quad
\delta^2R^0_i=-a^{-2}\delta^2R_{0i},
\end{equation}
\begin{equation}
\delta^2R^i_j=\delta^2g^{ik}\bar{R}_{kj}+\delta g^{ik}\delta R_{kj}+a^{-2}\delta^2R_{ij}.
\end{equation}
The Ricci scalar can be written as
\begin{equation}
\delta^2R=-a^{-2}\delta^2R_{00}+a^{-2}\delta^2 R_{ii}+\delta g^{ij}\delta R_{ij}+\delta^2g^{ij}\bar{R}_{ij},
\end{equation}
which leads to
\begin{equation}
a^2\delta^2 R=-\ao{\tilde{h}}^{ij}\ao{\tilde{h}}_{ij}''-\frac{3}{4}\ao{\tilde{h}}^{ij'}\ao{\tilde{h}}_{ij}'-3{\cal H}\ao{\tilde{h}}^{ij}\ao{\tilde{h}}_{ij}'+\ao{\tilde{h}}^{ij}\Delta \ao{\tilde{h}}_{ij}+\frac{3}{4}\ao{\tilde{h}}^{ij,k}\ao{\tilde{h}}_{ij,k}-\frac{1}{2}\ao{\tilde{h}}^{ij,k}\ao{\tilde{h}}_{ik,j}.
\end{equation}
The components of the Einstein tensor are 
\begin{equation}
-\frac{a^2}{2}\delta^2G^0_0=S_1,\label{00tensor}
\end{equation}
\begin{equation}
\quad \delta^2G^i_0=-\delta^2G^0_i=a^{-2}\delta^2R_{i0}
=\frac{2S_i}{a^2},\label{i0tensor}
\end{equation}
\begin{equation}
a^2\delta^2G^i_j=a^2(\delta^2g^{ik}\bar{R}_{kj}+\delta g^{ik}\delta R_{kj}+a^{-2}\delta^2R_{ij})=2S_3\delta_{ij}+2S_{ij}.\label{ijtensor}
\end{equation}
Eqs. (\ref{00scalar}) and (\ref{00tensor}) recover (\ref{timetime}) (see \umi{footnote\ref{sign}}).
Also, Eqs. (\ref{i0scalar}) and (\ref{i0tensor}) recover (\ref{timespace})\footnote{
\ao{The} indices "0" indicate $t$ in the previous subsection, while those indicate the conformal time $\eta$ in this subsection,
and they are related by $G^t_i=aG^\eta_i$.
}.
Let us decompose $S_{ij}$ as
$S_{ij}=S_4\delta_{ij}+S_{5,ij}+\cdots$, where $\cdots$ is to contain vector and tensor parts, which are irrelevant here. 
From this, we find $\Delta S^i_i=3\Delta S_4+\Delta^2 S_5$ 
and $S^{ij}_{\;\;,ij}=\Delta S_4+\Delta^2 S_5$, which lead to 
(\ref{S4}) and (\ref{S5}). Then we find (\ref{ein3}) and (\ref{ein4}) from (\ref{ijscalar1}), 
(\ref{ijscalar2}) and (\ref{ijtensor}).

The second-order parts of the divergence of the energy momentum tensor are
\begin{equation}
\delta^2T^\mu_{\nu;\mu}=\delta^2\Gamma^\mu_{\mu\lambda}\bar{T}^\lambda_\nu-\delta^2\Gamma^\lambda_{\mu\nu}\bar{T}^\mu_\lambda,
\end{equation}
which is nonzero when $\nu=0$:
\begin{equation}
\delta^2T^\mu_{0;\mu}=2(\rho+p)h^{ij}h_{ij}'.
\end{equation}
The negative of this gives the source term of (\ref{cons1}).
\section*{Appendix B: \blue{Derivation of\souta{ the expression of} the source term in Fourier space}}
In this appendix we derive (\ref{f1}) and (\ref{f2}). 
First, note that the Fourier components of $h^{ij}h_{ij}$ \umi{can be expressed as}
\be
(h^{ij}h_{ij})(\eta,\vec{k})=\int\fr{d^3\vec{k}'}{(2\pi)^{3/2}}\sum_{rs} h^r(\vec{k}')h^s(\vec{k}-\vec{k}')
D(\eta,k')E^{ij}_{rs\, ij}(\vec{k},\vec{k}') D(\eta,|\vec{k}-\vec{k}'|).
\ee
Similarly, the source can be written \souta{in the following form:} \sora{as}
\be
S(\eta,\vec{k})=\int\fr{d^3\vec{k}'}{(2\pi)^{3/2}}\sum_{rs}h^r(\vec{k}')h^s(\vec{k}-\vec{k}')
D(\eta,k')(\cdots)D(\eta,|\vec{k}-\vec{k}'|).\label{dots}
\ee
In the following, let us consider the contribution of each term \blue{in (\ref{source0})} to $(\cdots)$ of the above expression. 
The contribution of the term $\pa_jh_{ik}\pa^kh^{ij}=\pa_j\pa^k(h_{ik}h^{ij})$ in \blue{(\ref{S10})} to $(\cdots)$, indicated after the arrow in the equation below 
(the arrows elsewhere should be understood similarly), is 
\be
\pa_jh_{ik}\pa^kh^{ij}=\pa_j\pa^k(h_{ik}h^{ij})\quad\rightarrow\quad -k^2E^{rs}_1.
\ee
Similarly,
\be
\pa_kh_{ij}\pa^kh^{ij}=\fr{1}{2}\pa_k\pa^k(h_{ij}h^{ij})-(\Delta h_{ij})h^{ij}
\quad\rightarrow\quad
-\fr{1}{2}k^2E^{rs}_2+k^{'2}E^{rs}_2.
\ee
So the contribution of $S_1$ is 
\be
S_1
\quad\rightarrow\quad
\left(
-\fr{1}{4}\overleftarrow{\pa_\eta}\pa_\eta+\pa_\eta^2-\fr{3}{8}k^2+\fr{3}{4}k^{'2}
\right)E^{rs}_2
+\fr{k^2}{2}E^{rs}_1,\label{S1}
\ee
where $\overleftarrow{\pa_\eta}$ is supposed to differentiate \textit{only} $D(\eta,k')$ of Eq.(\ref{dots}) in the left.
Likewise, the contribution of $S_{3}$ is 
\be
S_{3}
\quad\rightarrow\quad
\left(
\fr{3}{4}\overleftarrow{\pa_\eta}\pa_\eta+\fr{3}{8}k^2-\fr{3}{4}k^{'2}
\right)E^{rs}_2
-\fr{k^2}{2}E^{rs}_1.
\ee
To obtain the contribution of $\kh^i\kh^jS_{ij}$, let us rewrite $S_{ij}$ as follows:
\begin{align}
S_{ij}=-h_i^{~k'}h_{jk}'+\pa_k\pa_l(h^{kl}h_{ij})-\pa_l(h^{kl}\pa_ih_{jk})-(i\leftrightarrow j)
-\pa_k\pa^l(h_{jl}h_i^{~k})\nonumber\\
+\pa_lh_{jk}\pa^lh_i^{~k}+
\fr{1}{2}\pa_i\pa_j(h^{kl}h_{kl})
-\fr{1}{2}\pa_ih^{kl}\pa_jh_{kl}.
\end{align}
Then, the contribution is 
\begin{align}
S_{ij}\kh^i\kh^j
\quad\rightarrow\quad
-\overleftarrow{\pa_\eta}\pa_\eta E_{i~jk}^{~k}\kh^i\kh^j-k_kk_lE^{kl}_{~~ij}\kh^i\kh^j
+2k_l(k_i-k_i')E^{kl}_{~~jk}\kh^i\kh^j\nonumber\\
+k_kk^lE_{jli}^{~~k}\kh^i\kh^j
-k_l'(k^l-k^{'l})E_{jki}^{~~~k}\kh^i\kh^j
-\fr{1}{2}k_ik_j\kh^i\kh^jE^{kl}_{~~kl}\nonumber\\
+\fr{1}{2}k'_i(k_j-k_j')E^{kl}_{~~kl}\kh^i\kh^j\nonumber\\
=(
-\overleftarrow{\pa_\eta}\pa_\eta+2k^2-3kk'\mu+k^{'2}
)E^{rs}_1
+\fr{1}{2}(k'\mu(k-k'\mu)-k^2)E^{rs}_2.
\end{align}
The collection of all the contributions yields
\begin{align}
&S
\quad\rightarrow\quad
\left\{
\overleftarrow{\pa_\eta}\pa_\eta-\fr{1}{2}(3-c_{\rm{s}}^2)k^2+3kk'\mu-k^{'2}
\right\}E^{rs}_1+\nonumber\\
&\left\{
-\fr{1}{4}(3+c_{\rm{s}}^2)\overleftarrow{\pa_\eta}\pa_\eta+c_{\rm{s}}^2\pa_\eta^2+2c_{\rm{s}}^2\H \pa_\eta
+\fr{1}{8}(1-3c_{\rm{s}}^2)k^2-\fr{1}{2}k'\mu(k-k'\mu)+\fr{3}{4}(1+c_{\rm{s}}^2)k^{'2}
\right\}E_2^{rs},\label{contribution}
\end{align}
from which (\ref{f1}) and (\ref{f2}) can be read off. 
\section*{Appendix C: \blue{\souta{Numerical calculation of the }PDF of \sora{induced} radiation density perturbations}}
In this paper \souta{${\cal P}_h(k)$ is assumed to be } \umi{a delta-function\sora{-}type} \sora{tensor spectrum is assumed} (see (\ref{Ph}))\sora{, but since it cannot be treated in}\souta{ \ao{However}, since a delta function can not be used in} numerical calculations adopting discretization in Fourier space, the power spectrum is instead approximated by the following top-hat \souta{\umi{shape}} \sora{spectrum} \souta{in the simulation:} \sora{here:}
\be
{\cal P}_h(k)={\cal A}^2\epsilon^{-1}\:\left(k_p\left[1-\fr{\epsilon}{2}\right]<k<k_p\left[1+\fr{\epsilon}{2}\right]\right), \quad 0\:(\mathrm{otherwise}). 
\ee
In this appendix, we set ${\cal A}=1$, 
\blue{and $\epsilon$ is \souta{to be set to sufficiently small values} \sora{chosen to be sufficiently small}, as presented shortly}.

\begin{figure}[t]
\begin{center}
\includegraphics[width=7cm,keepaspectratio,clip]{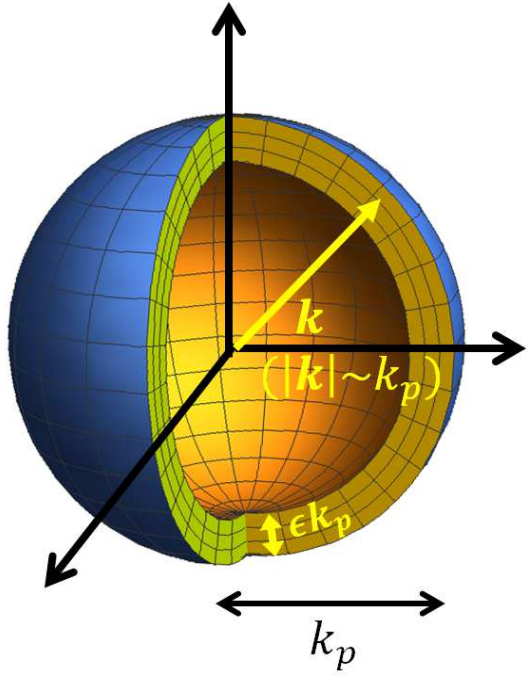}
\end{center}
\caption{\blue{An illustration of the spherical shell in the Fourier space 
considered in this appendix.} 
}
\label{sphere}
\end{figure}

Let us decompose the Fourier components of $h^{r}(\vec{k})$ as follows:
\be
h^{r}(\vec{k})=a^r(\vec{k})+ib^r(\vec{k}),
\ee
where $a^r$ and $b^r$ are real Gaussian random variables satisfying
\be
a^+(-\vec{k})=a^+(\vec{k}),\quad b^+(-\vec{k})=-b^+(\vec{k}),\quad a^{\times}(-\vec{k})=-a^{\times}(\vec{k}),\quad b^{\times}(-\vec{k})=b^{\times}(\vec{k})
\ee
to ensure the reality of $h_{ij}(\eta,\vec{x})$ (note that $e_{ij}^{\times}(-\vec{k})=-e_{ij}^{\times}(\vec{k})$ as well as $e_{ij}^{+}(-\vec{k})=e_{ij}^{+}(\vec{k})$ following the definitions of \sora{the} polarization 
tensors we adopt). 
\souta{In Monte Carlo simulations, 
w} \sora{W}e consider a spherical shell in the Fourier space whose radius is $k_p$ and whose \umi{thickness} is $\epsilon k_p$, \blue{as is depicted in Fig. \ref{sphere}}. 
Let us \ao{denote} the grid points in this \blue{spherical shell} by $\vec{k}_i$, where $i$ is a natural number.
Each \blue{of these grid points} is associated with two complex numbers $h^{r}(\vec{k}_i)=a^r(\vec{k}_i)+ib^r(\vec{k}_i), (r=+,\times)$ (satisfying $h^+(-\vec{k}_i)=h^{+}(\vec{k}_i)^*, 
h^{\times}(-\vec{k}_i)=-h^{\times}(\vec{k}_i)^*$), 
where the dispersion of both $a^r$ and $b^r$ is \souta{given by} 
\be
\sigma^2=\fr{\pi^2}{k_p^3}dk^{-3}\epsilon^{-1},
\ee
with $dk$ denoting the interval between two neighboring grid points in the Fourier space. 
Then, \umi{from (\ref{deltar})}\sora{,} $\delta_r(\eta,\vec{x}=0,R)$ for a specific realization of $\{h^r(\vec{k}_i)\}$ is calculated by 
\begin{align}
\delta_r(\eta,\vec{x}=0,R)=
\fr{1+c_{\rm{s}}^2}{c_{\rm{s}}^2{\cal H}}
\fr{(dk)^6}{(2\pi)^3}\biggl\{\sum_{r,s}\sum_{\green{\vec{{\scriptstyle k}}_i,\vec{{\scriptstyle k}}_j\in S}}W(\green{|\vec{k}_i+\vec{k}_j|}R)h^{r}(\vec{k}_i)h^{s}(\vec{k}_j)F_{rs}(\eta,\green{\vec{k}_i+\vec{k}_j,\vec{k}_i})\biggr\},\label{simulation}
\end{align}
where $S$ denotes the set comprised of the grid points inside the spherical shell.
As mentioned in the main text, we set $\eta=R=(\green{0.7}k_p)^{-1}$. 
\ao{
When one is interested in the power spectrum, (\ref{E1}) and (\ref{E2}) can be used due to isotropy, 
but in simulations 
their components, including nonzero cross terms,
for each combination of wave vectors have to be explicitly calculated using
}
\be
e_{ij}^{+}(\hat{k})=
\left(
\begin{array}{ccc}
1 & 0&0  \\
 0& -1&0  \\
 0&0 &0 
\end{array}
\right)(\green{|\hat{k}_3|}=1)
,\quad\left(
\begin{array}{ccc}
\frac{\hat{k}_3^2-\hat{k}_3^4-\hat{k}_2^2(1+\hat{k}_3^2)}{-1+\hat{k}_3^2} &\frac{\hat{k}_1\hat{k}_2(1+\hat{k}_3^2)}{-1+\hat{k}_3^2} &\hat{k}_1\hat{k}_3  \\
\frac{\hat{k}_1\hat{k}_2(1+\hat{k}_3^2)}{-1+\hat{k}_3^2} &\frac{-1+\hat{k}_3^2+\hat{k}_2^2(1+\hat{k}_3^2)}{-1+\hat{k}_3^2} &\hat{k}_2\hat{k}_3  \\
\hat{k}_1\hat{k}_3 &\hat{k}_2\hat{k}_3 &-1+\hat{k}_3^2 
\end{array}
\right)(\green{|\hat{k}_3|}\neq 1),
\ee
\be
e_{ij}^{\times}(\hat{k})=\pm
\left(
\begin{array}{ccc}
0 & 1&0  \\
 1& 0&0  \\
 0&0 &0 
\end{array}
\right)(\hat{k}_3=\pm 1)
,\quad\left(
\begin{array}{ccc}
-\frac{2\hat{k}_1\hat{k}_2\hat{k}_3}{-1+\hat{k}_3^2} &-\frac{\hat{k}_3(-1+2\hat{k}_2^2+\hat{k}_3^2)}{-1+\hat{k}_3^2} &-\hat{k}_2  \\
-\frac{\hat{k}_3(-1+2\hat{k}_2^2+\hat{k}_3^2)}{-1+\hat{k}_3^2} &\frac{2\hat{k}_1\hat{k}_2\hat{k}_3}{-1+\hat{k}_3^2} &\hat{k}_1  \\
-\hat{k}_2 &\hat{k}_1 &0 
\end{array}
\right)(\green{|\hat{k}_3|}\neq 1).
\ee

\umi{\souta{With the aid of}\sora{Using} some of its symmetry properties, (\ref{simulation}) can be simplified\souta{, facilitating numerical calculations,} as follows.}
\green{
Let us denote by $S/2$ the set of the grid points inside the upper half of the spherical shell\souta{ $S$}.
More precisely, the set $S/2$ is made up of the grid points $\nami{\vec{k}_i}$ in the spherical shell\souta{ $S$} 
with $(\vec{k}_i)_z>0$, those with $(\vec{k}_i)_z=0$ and $(\vec{k}_i)_y>0$, and also those with $(\vec{k}_i)_z=(\vec{k}_i)_y=0$
and $\ao{(\vec{k}_i)_x}>0$.
Then, the inside of the brace of (\ref{simulation}) can be rewritten as 
\begin{align}
&\sum_{r,s}\sum_{\vec{{\scriptstyle k}}_i,\vec{{\scriptstyle k}}_j\in S/2}
\kaku{W(|\vec{k}_i+\vec{k}_j|)\nami{h^r(\vec{k}_i)h^s(\vec{k}_j)
F_{rs}(\eta,\vec{k}_i+\vec{k}_j,\vec{k}_i)
+h^r(-\vec{k}_i)h^s(-\vec{k}_j)
F_{rs}(\eta,-\vec{k}_i-\vec{k}_j,-\vec{k}_i)}\right.\nonumber\\
&\left.\qquad\qquad\qquad+
W(|\vec{k}_i-\vec{k}_j|)\nami{h^r(-\vec{k}_i)h^s(\vec{k}_j)
F_{rs}(\eta,-\vec{k}_i+\vec{k}_j,-\vec{k}_i)
+h^r(\vec{k}_i)h^s(-\vec{k}_j)
F_{rs}(\eta,\vec{k}_i-\vec{k}_j,\vec{k}_i)}
}\nonumber\\ \nonumber\\
&=\sum_{r,s}\sum_{\vec{{\scriptstyle k}}_i,\vec{{\scriptstyle k}}_j\in S/2}
\kaku{W(|\vec{k}_i+\vec{k}_j|)\nami{h^r(\vec{k}_i)h^s(\vec{k}_j)
+h^r(\vec{k}_i)^*h^s(\vec{k}_j)^*}
F_{rs}(\eta,\vec{k}_i+\vec{k}_j,\vec{k}_i)
\right.\nonumber\\
&\left.\qquad\qquad\qquad\quad+
W(|\vec{k}_i-\vec{k}_j|)\epsilon_s\nami{
h^r(\vec{k}_i)^*h^s(\vec{k}_j)+h^r(\vec{k}_i)h^s(\vec{k}_j)^*}
F_{rs}(\eta,\vec{k}_i-\vec{k}_j,\vec{k}_i)
}\nonumber\\ \nonumber\\
&=\sum_{r,s}\sum_{\vec{{\scriptstyle k}}_i,\vec{{\scriptstyle k}}_j\in S/2}
\kaku{2W(|\vec{k}_i+\vec{k}_j|)\nami{a^r(\vec{k}_i)a^s(\vec{k}_j)
-b^r(\vec{k}_i)b^s(\vec{k}_j)}
F_{rs}(\eta,\vec{k}_i+\vec{k}_j,\vec{k}_i)
\right.\nonumber\\ 
&\left.\qquad\qquad\qquad\quad+
2\epsilon_sW(|\vec{k}_i-\vec{k}_j|)\nami{
a^r(\vec{k}_i)a^s(\vec{k}_j)+b^r(\vec{k}_i)b^s(\vec{k}_j)}
F_{rs}(\eta,\vec{k}_i-\vec{k}_j,\vec{k}_i)
},\label{brace}
\end{align}
where we have used $h^r(-\vec{k}_i)=\epsilon_rh^r(\vec{k}_i)^* \;(\epsilon_+=1,\;\epsilon_\times~-1)$ and 
$F_{rs}(\eta,-\vec{k},-\vec{k}')=\epsilon_r\epsilon_sF_{rs}(\eta,\vec{k},\vec{k}')$. \ao{This} has explicitly proven that $\delta_r$ is real, as it should. 
Let us label the grid points in $S/2$ by $1,2,\cdots,N$, then introducing 
\begin{equation}
\vec{a}^t=\sigma^{-1}(a^+(\vec{k}_1),a^+(\vec{k}_2),\cdots,a^+(\vec{k}_N),a^\times(\vec{k}_1),a^\times(\vec{k}_2),\cdots,a^\times(\vec{k}_N)),
\end{equation}
\begin{equation}
\vec{b}^t=\sigma^{-1}(b^+(\vec{k}_1),b^+(\vec{k}_2),\cdots,b^+(\vec{k}_N),b^\times(\vec{k}_1),b^\times(\vec{k}_2),\cdots,b^\times(\vec{k}_N)),
\end{equation}
and using (\ref{brace}) we can rewrite (\ref{simulation}) as 
\begin{equation}
\delta_r(\eta,\vec{x}=\vec{0},R)=\f{1+c_s^2}{c_s^2{\cal H}}\f{dk^3}{8\pi\epsilon k_p^3}\nami{\vec{a}^t
\vec{M}^a
\vec{a}+\vec{b}^t
  \vec{M}^b
\vec{b}},\label{matrix}
\end{equation}
where
\begin{equation}
\vec{M}^a\equiv
\left(
    \begin{array}{cc}
      \vec{M}_{++}^a & \vec{M}^a_{+\times}  \\
      \vec{M}^a_{\times+} & \vec{M}^a_{\times\times} 
    \end{array}
  \right),\quad
  \vec{M}^b\equiv
  \left(
    \begin{array}{cc}
      \vec{M}_{++}^b & \vec{M}^b_{+\times}  \\
      \vec{M}^b_{\times+} & \vec{M}^b_{\times\times} 
    \end{array}
  \right),
\end{equation}
\begin{equation}
(\vec{M}_{rs}^a)_{ij}=(\vec{M}^1_{rs})_{ij}+(\vec{M}^2_{rs})_{ij},
\end{equation}
\begin{equation}
(\vec{M}_{rs}^b)_{ij}=-(\vec{M}^1_{rs})_{ij}+(\vec{M}^2_{rs})_{ij},
\end{equation}
\begin{equation}
(\vec{M}^1_{rs})_{ij}=2W(|\vec{k}_i+\vec{k}_j|)F_{rs}(\eta,\vec{k}_i+\vec{k}_j,\vec{k}_i),
\end{equation}
\begin{equation}
(\vec{M}^2_{rs})_{ij}=2\epsilon_sW(|\vec{k}_i-\vec{k}_j|)
F_{rs}(\eta,\vec{k}_i-\vec{k}_j,\vec{k}_i).
\end{equation}
Noting $F_{rs}(\eta,\vec{k}_i+\vec{k}_j,\vec{k}_i)=F_{sr}(\eta,\vec{k}_j+\vec{k}_i,\vec{k}_j)$ and 
$\epsilon_sF_{rs}(\eta,\vec{k}_i-\vec{k}_j,\vec{k}_i)=\epsilon_rF_{sr}(\eta,\vec{k}_j-\vec{k}_i,\vec{k}_j)$,
one can confirm \sora{that} $\vec{M}^a$ and $\vec{M}^b$ are symmetric matrices. 
So by diagonalizing $\vec{M}^a$ and $\vec{M}^b$ \souta{one can further rewrite }(\ref{matrix}) \sora{can be further rewritten} as 
}
\be
\delta_r=a_0\sum_{i=1}^{\green{2}N}a_ix_i^2,\label{diagonal}
\ee
where $x_1,x_2,\cdots$ are independent Gaussian random variables whose dispersion is unity and 
\be
0<a_0,\quad 1=|a_1|>|a_2|>\cdots>|a_{4N}|.
\ee
\green{
\souta{The}\sora{Its} average and dispersion\souta{ of $\delta_r$} \souta{can be written as follows:} \sora{are}
\begin{equation}
\sankaku{\delta_r}=a_0\sum_{i=1}^{2N}a_i,
\end{equation}
\begin{equation}
\sigma^2=\langle\delta_r^2\rangle-\sankaku{\delta_r}^2
=a_0^2\maru{\sum_{i=1}^{2N}a_i^2\sankaku{x_i^4}+\sum_{i\neq j}a_ia_j-\sum_{i=1}^{2N}a_i^2-\sum_{i\neq j}a_ia_j}
=2a_0^2\sum_{i=1}^{2N}a_i^2.
\end{equation}
These \souta{quantities } can also be calculated from Eqs. (\ref{average}) and (\ref{rootmean}). For $\eta=(0.7k_p)^{-1}$, 
$\sankaku{\delta_r}\simeq -0.69$ and $\sigma\simeq 1.03$, and these values have also been obtained in the numerical computations 
(see (\ref{coe}) below), which serves as a crosscheck.
We chose $\epsilon=0.05$ and $dk=\epsilon k_p$. In this case, $N$ turns out to be 2517, but interestingly more than
95\% of $\delta_r$ is determined by only the first 24 terms with \souta{most of } the rest\souta{ of the terms} \souta{being vanishingly small} \sora{negligible}, namely, 
\begin{equation}
\frac{\sum_{i=1}^{24}|a_i|}{\sum_{i=1}^{2N}|a_i|}\blue{\simeq 0.98}.
\end{equation}
\souta{Therefore}\sora{Consequently}, in \souta{analyzing} \sora{deriving} the PDF of $\delta_r$ one can \souta{just } focus only on \souta{a limited number of terms} \sora{them}, safely neglecting \souta{most of the terms} \sora{the rest}, 
which greatly simplifies the analysis. We found
\begin{align}
&\sankaku{\delta_r}\simeq-0.69,\;\sigma\simeq1.0,\;a_0\simeq0.30,\;a_{1-5}\simeq-1.0,\nonumber\\
&\;a_{6-10}\simeq0.46,\;a_{11-17}\simeq0.11,\;a_{18-24}\simeq-0.078.\label{coe}
\end{align}
We have also calculated the coefficients for $\epsilon=0.1$,$dk=\epsilon k_p$ and also for $\epsilon=0.1$,$dk=2\epsilon k_p/3$, 
\sora{and} the results \souta{of which } coincided with the above well.
\souta{This indicates} \sora{Hence we can conclude} that the above choices of $\epsilon=0.05$ and $dk=\epsilon k_p$ are sufficiently small to obtain reliable results. 
}

\sora{We are in a position to} discuss the PDF of $\delta_r$ using the coefficients of (\ref{coe}). 
First one can resort to a brute-force method of a Monte Carlo simulation to obtain the PDF of $\delta_r$, 
by simply generating 2\green{4} random Gaussian variables with dispersion unity, ${x_1,x_2,\cdots, x_{2\green{4}}}$, and summing up the square of them with the coefficients \blue{above}. We have generated $\{x_i\}$ \blue{a million} times to \souta{\ao{construct}} \sora{obtain} the PDF of $\delta_r$, \souta{with the result } shown in Fig. \ref{pdf}. 
\green{In this appendix ${\cal A}$ is set to unity, and so what is shown there is the PDF of $\tilde{\delta_r}\equiv(\delta_r-\langle\delta_r\rangle)/{\cal A}^2$.}

\green{
We adopt \sora{the} Clopper-Pearson interval \cite{Clopper} to obtain \sora{the}
95\% confidence interval $p_L<p<p_U$ of the probability $p$
of $\delta_r$ being realized in some interval $(\delta_r\pm d\delta_r)$, when $\delta_r$ 
in that range is realized $k$ times in $N$ trials\sora{, as follows}. 
First, the number of an event with probability 
$p$ realized in $N$ trials follows a Binomial distribution:
$P(k;p)={}_NC_kp^k(1-p)^{N-k}$. 
Let us introduce $\alpha=1-C, C=0.95.$
From the meaning of the confidence interval, the probability 
of the event \souta{begin} \sora{being} realized less than $k$ times when $p=p_U$ 
is $\alpha/2:$
\begin{equation}
\sum_{i=0}^kP(i;p_U)=I(1-p_U,N-k,1+k)
=1-I(p_U,1+k,N-k)=\frac{\alpha}{2},
\end{equation}
where $I(x,a,b)$ is the regularized beta function and the relation 
$I(x,a,b)=I(1-x,b,a)$ has been used. From this, 
$p_U$ can be expressed by the inverse $I^{-1}$ of the regularized beta function as
\begin{equation}
p_U=I^{-1}\left(1-\frac{\alpha}{2},1+k,N-k\right).\label{pU}
\end{equation}
Similarly, the probability of the event being realized more than 
$k$ times when $p=p_L$ is $\alpha/2$:
\begin{equation}
\sum_{i=k}^NP(i;p_L)=1-I(1-p_L,N-k+1,k)
=I(p_L,k,N-k+1)=\frac{\alpha}{2},
\end{equation}
which leads to 
\begin{equation}
p_L=I^{-1}\left(\frac{\alpha}{2},k,N-k+1\right).\label{pL}
\end{equation}
The error bars in Fig. \ref{pdf} are obtained from (\ref{pU}) and (\ref{pL}).
}

\umi{Finally} let us discuss \umi{an} approximate \souta{form of} \sora{formula for} the PDF. 
Noting \sora{that} the first ten terms of (\ref{diagonal}) give dominant contributions, 
\souta{let us} \sora{we} \umi{begin by \souta{considering}} \sora{deriving} the PDF of \souta{the form } $Z=-X+cY$, where $X$ and $Y$ are both random variables following \umi{the} chi-squared distribution with $n$ degrees of freedom and $c$ is a positive constant. 
The PDF of both X and Y is\souta{ given by} 
\be
P_1(n;X)=\fr{(1/2)^{n/2}}{\Gamma(n/2)}X^{n/2-1}e^{-X/2}.
\ee
Then the PDF of $Z$ is\souta{ calculated as follows:}
\begin{align}
P_2(n,c;Z)&=\blue{N_1}\int_0^\infty dX\int_0^\infty dY\delta(Z+X-cY)P_1(\umi{n;}X)P_1(\umi{n;}Y)\nonumber\\
&=\frac{\blue{N_1}(1/2)^n}{\Gamma(n/2)^2}e^{-\fr{Z}{2c}}\left(\frac{1}{c}\right)^{\fr{n}{2}-1}\int_{\max\{0,-Z\}}^\infty dX X^{\fr{n}{2}-1}e^{-\fr{X}{2}}(Z+X)^{\fr{n}{2}-1}e^{-\fr{X}{2c}}\nonumber\\
&=\frac{\blue{N_1}}{\sqrt{2\pi 2^n}\Gamma(n/2)}c^{1-n/2}\exp\left(-\frac{1-c}{4c}Z\right)
\left(\frac{c|Z|}{1+c}\right)^{(n-1)/2}K_{(n-1)/2}\left(\frac{1+c}{4c}|Z|\right),
\end{align}
where $\blue{N_1}$ is a normalization factor and $K_m(x)$ is the modified Bessel function of second kind. 
In \souta{considering} \sora{deriving} the PDF of $\delta_r$, one may simply replace the terms $11\leq i$ in (\ref{diagonal}) 
by their expectation values $E\equiv 7a_{11}+\green{7}a_{18}$\souta{, noting the relative unimportance of these terms} \sora{since they are relatively unimportant}, and then \sora{finally} \souta{the PDF of $\delta_r$ is finally given by}
\be
P\left(\tilde{\delta}_r\right)\umi{\simeq}\:P_2\left(5,a_6;\frac{\tilde{\delta}_r+\langle\delta_r\rangle/{\cal A}^2}{a_0}-E\right).\label{last}
\ee 
Interestingly, this approximates the PDF inferred from the Monte Carlo simulation mentioned above overall fairly well, as is shown in Fig. \ref{pdf}. 
\green{
In more detail, this formula slightly deviates from the simulated points around $\tilde{\delta}_r\sim 0$, presumably because 
the terms $11\leq i$, simply replaced by their expectation values to obtain the above approximate formula, are relatively important \souta{in this region} \sora{there}. 
On the other hand, \souta{the approximate} \sora{this} formula is better for $|\tilde{\delta}_r|\gtrsim 2,$ which is probably because the probability of 
these relatively rare events is mostly determined by the first ten terms, with the rest of the terms lying around their expectation 
values. Since the probability of PBH formation has to be extremely rare, what matters is only the \souta{positive tail part} \sora{tail} of the PDF, 
and \souta{so} \sora{therefore} we can safely use \souta{this} \sora{the above} approximate formula to calculate \umi{the} PBH\sora{s'} abundance and place upper bounds on tensor perturbations 
from \souta{the absence of PBHs} \sora{their absences}. 
}

\section*{ACKNOWLEDGMENTS}
We are grateful to Jun'ichi Yokoyama for reading the manuscript, useful comments and continuous encouragement.
We also thank Kazunari Eda, Yuki Watanabe, Yosuke Itoh, Daisuke Yamauchi, 
Tsutomu Kobayashi, Masahide Yamaguchi, Takahiro Tanaka for helpful comments.
This work was partially
supported by Grant-in-Aid for JSPS Fellow No. 25.8199 (T.N.),
JSPS Postdoctoral Fellowships for Research Abroad (T.N.), 
\ao{JSPS Grant-in-Aid for Young Scientists (B) No. 15K17632 (T.S.),
MEXT Grant-in-Aid for Scientific Research on Innovative Areas 
``New Developments in Astrophysics Through Multi-Messenger 
Observations of Gravitational Wave Sources'' No. 15H00777 (T.S.) and 
``Cosmic Acceleration'' No. 15H05888 (T.S.)}

\bibliographystyle{h-physrev}
\bibliography{ref}

\end{document}